\documentclass[11pt]{article}

\usepackage[T1]{fontenc}
\usepackage[utf8]{inputenc}
\usepackage{lmodern}

\usepackage{mathtools, amsthm}

\usepackage{isomath} % ISO standard math setup
\newcommand*{\+}[1]{\ensuremath{\vectorsym{#1}}}  % ISO bold italic math symbol

\usepackage{dsfont}

\RequirePackage[pdfusetitle,colorlinks,citecolor=blue,urlcolor=blue]{hyperref}

\RequirePackage{natbib}
\bibpunct{(}{)}{;}{a}{,}{,}

\usepackage[nottoc]{tocbibind}
\settocbibname{References}

\usepackage[page,titletoc]{appendix}

\usepackage[format=plain, font=it]{caption}

\usepackage{subcaption}

\usepackage[all,poly,ps,color]{xy}
\usepackage{url,epsfig,amssymb}
\usepackage{color}

\usepackage{framed}
\usepackage{microtype}

\usepackage{geometry}
\usepackage{sectsty}
\usepackage[normalem]{ulem}
\sectionfont{\sffamily\bfseries\upshape\large}
\subsectionfont{\sffamily\bfseries\upshape\normalsize}
\subsubsectionfont{\sffamily\bfseries\upshape\normalsize}
\makeatletter

\renewcommand\@seccntformat[1]{\csname the#1\endcsname.\quad}

\makeatletter
\def\@maketitle{%
  \begin{center}%
  \let \footnote \thanks
    {\large \@title \par}%
    {\normalsize
      \begin{tabular}[t]{c}%
        \@author
      \end{tabular}\par}%
    {\small \@date}%
  \end{center}%
}
\makeatother

\newenvironment{example}[1]{\begin{quote}{\bf Example.  #1}\\}{\end{quote}}

\newcommand{\newc}{\newcommand}
\newc{\reals}{\ensuremath{\mathbb{R}}}
\newc{\nnb}{\nonumber}
\newc{\beqann}{\begin{eqnarray*}}
\newc{\beqa}{\begin{eqnarray}}
\newc{\beqnn}{\begin{displaymath}}
\newc{\beq}{\begin{equation}}
\newc{\bex}{\begin{example}}
\newc{\eex}{\end{example}}
\newc{\eeqann}{\end{eqnarray*}}
\newc{\eeqa}{\end{eqnarray}}
\newc{\eeqnn}{\end{displaymath}}
\newc{\eeq}{\end{equation}}
\newc{\E}{\mbox{E}}
\newc{\Invc}{\mbox{Inv-$\chi^2$}}       %{\mbox{Inverse-$\chi^2$}}
\newc{\Nor}{\mbox{N}}
\newc{\Ber}{\mbox{Bernoulli}}
\newc{\Var}{\mbox{var}}
\newc{\btab}{\begin{tabular}}
\newc{\etab}{\end{tabular}}
\newc{\rep}{\rm rep}

\usepackage{ragged2e}

\usepackage{pgf}
\usepackage{pgfplots}
\usetikzlibrary{bayesnet}

\let\originalleft\left
\let\originalright\right
\renewcommand{\left}{\mathopen{}\mathclose\bgroup\originalleft}
\renewcommand{\right}{\aftergroup\egroup\originalright}

\usepackage{gensymb}

\title{\bf Expectation propagation as a way of life:\\
A framework for Bayesian inference on partitioned data\vspace{\baselineskip}}

\author{
Aki Vehtari\thanks{Department of Computer Science, Aalto University, Finland.}
\and
Andrew Gelman\thanks{Department of Statistics, Columbia University, New York.}
\and
Tuomas Sivula\footnotemark[1]
\and
Pasi Jyl{\"a}nki\thanks{Donders Institute for Brain, Cognition, and Behavior, Radboud University Nijmegen, Netherlands.}
\and
Dustin Tran\footnotemark[2]
\and
Swupnil Sahai\footnotemark[2]
\and
Paul Blomstedt\footnotemark[1]
\and
John~P. Cunningham\footnotemark[2]
\and
David Schiminovich\thanks{Department of Astronomy, Columbia University, New York.}
\and
Christian Robert\thanks{Universit{\'e} Paris Dauphine.}
}

\date{}

\makeatletter
\hypersetup{
  pdftitle={\@title},
  pdfauthor={%
    Aki Vehtari,
    Andrew Gelman,
    Tuomas Sivula,
    Pasi Jyl{\"a}nki,
    Dustin Tran,
    Swupnil Sahai,
    Paul Blomstedt,
    John~P. Cunningham,
    David Schiminovich,
    Christian Robert},
  pdfkeywords={%
    Bayesian computation,
    big data,
    data partitioning,
    expectation propagation,
    hierarchical models,
    statistical computing}
}
\makeatother

\begin{document}

\maketitle
\thispagestyle{empty}

\begin{abstract}

A common divide-and-conquer approach for Bayesian computation with big data is to partition the data, perform local inference for each piece separately, and combine the results to obtain a global posterior approximation.
While being conceptually and computationally appealing, this method involves the problematic need to also split the prior for the local inferences; these weakened priors may not provide enough regularization for each separate computation, thus eliminating one of the key advantages of Bayesian methods.
To resolve this dilemma while still retaining the generalizability of the underlying local inference method, we apply the idea of expectation propagation (EP) as a framework for distributed Bayesian inference.
The central idea is to iteratively update approximations to the local likelihoods given the state of the other approximations and the prior.

The present paper has two roles: we review the steps that are needed to keep EP algorithms numerically stable, and we suggest a general approach, inspired by EP, for approaching data partitioning problems in a way that achieves the computational benefits of parallelism while allowing each local update to make use of relevant information from the other sites.
In addition, we demonstrate how the method can be applied in a hierarchical context to make use of partitioning of both data and parameters.
The paper describes a general algorithmic framework, rather than a specific algorithm, and presents an example implementation for it.

\end{abstract}

\section{Introduction}
\label{sec:introduction}

Expectation propagation (EP) is a fast and parallelizable method of
distributional approximation via data partitioning.
Since its introduction by \citet{Opper+Winther:2000} and
\citet{Minka:2001b}, EP has been an important
Bayesian computational method for inferring intractable posterior
densities.

Motivated by the substantial methodological progress made in the last decade and a half,
our aim in this paper is to review the current state of the art,
also serving readers with no previous exposure to EP as an introduction to the methodology.
The main theme of our paper is to use EPs message passing technique as a framework for distributed Bayesian inference.
The problem is treated in the general setting of combining inferences on data partitioned into disjoint subsets.
This setting can be motivated from two complementary views of distributed computing, top-down and bottom-up,
both of which have gained increasing attention in the statistics and machine learning communities.
We approach them as instances of the same computational framework.

The top-down view deals with fitting statistical models to large
data sets, for which many distributed (divide-and-conquer) algorithms have been proposed over the
past few years~\citep{Ahn+others:2012,Korattikara+others:2014,Hoffman+others:2013,Scott+others:2016,Wang+Dunson:2013,Neiswanger+others:2014}.
The motivation for distributing the inference may be to decrease run time or  deal with memory limitations.
The basic idea is to partition the data $y$ into $K$ pieces, $y_1,\dots,y_K$, each with
likelihood $p(y_k|\theta)$, then analyze each part of the likelihood separately, and finally combine the
$K$ pieces to perform inference (typically approximately) for $\theta$.

In a Bayesian context, though, it is not clear how distributed computations should handle the prior distribution.
If the prior $p(\theta)$ is included in each separate inference, it will be multiply counted when the $K$
inferences are combined. To correct for this, one can in principle divide the combined posterior by $p(\theta)^{K-1}$ at the end, but this can lead to computational instabilities. An alternative is to
divide the prior itself into pieces, but then the fractional prior $p(\theta)^{1/K}$ used for each
separate computation may be too weak to effectively regularize, thus eliminating one of the key
computational advantages of Bayesian inference; for examples in which the likelihood alone does not allow
good estimation of $\theta$, see \citet{Gelman+others:1996}, \citet{Gelman+others:2008}, and, in the
likelihood-free context, \citet{Barthelme+Chopin:2014}.

Turning to the bottom-up view, the motivation for distributed inference may come from the local nature of the data and the model.
Here the data---not necessarily big in size---are already split into $K$
pieces, each with likelihood $p(y_k|\theta)$.
For example, in privacy-preserving computing, the data owners of local pieces can
only release aggregated
information such
as moments~\citep[e.g.][]{Sarwate+others:2014,Dwork+Roth:2014}. In meta-analysis,
different pieces of information come from different sources or are
reported in different ways, and the task is to combine
such information~\citep{Dominici+others:1999,Higgins+Whitehead:1996}.
In both settings, we
would like to partially pool across separate analyses,
enabling more informed decisions both globally and for the local analyses.
These types of problems fall into the general framework of hierarchical
models, and---as in the privacy-preserving setting---may need to be
solved without any single processor having complete access to all the local data or model.

%The EP idea
Extracting the core principles behind EP motivates a general framework for passing information between inferences on partitioned data.
In classical EP, the data are typically partitioned pointwise, with the approximating density fully factorized.
When data are partitioned into bigger subsets, the same idea can be used in a more versatile manner.
Here the  \emph{cavity distribution}, which approximates the effect of inferences from all other $K-1$ data partitions, can be used as a prior in the inference step for individual partitions.
\citet{Xu+others:2014} apply the EP algorithm in a distributed setting by using MCMC for performing inference for the separate sites.
A posterior server maintains a global approximation and iteratively issues updates for distributed sites. At each site, a local inference is carried out and the obtained posterior sample is used to form moment estimates for updating the local approximation. The site updates are propagated back to the posterior server and the global approximation is updated.
Motivated by an earlier preprint version of the present paper, \citet{hernandez_hernandez_2016} apply a similar distributed approach for Gaussian process classification. Instead of MCMC, they apply nested EP updates in each site and use the same distributed cluster for updating the hyperparameters between iterations.

EP is not in general guaranteed to converge, which motivates an alternative direction of algorithmic development that, instead of local updates, applies various energy optimization techniques directly to the related objective function.
Motivated by \citet{Xu+others:2014} and our earlier preprint version of the present paper \citep{Gelman+others:2014b}, \citet{Hasenclever+others:2017} develop such a distributed algorithm called stochastic natural gradient expectation propagation (SNEP), which also uses MCMC for the site inferences.
Similar to the methods presented by~\citet{Heskes+Zoeter:2002} and~\citet{Opper+Winther:2005}, they implement a convergent double-loop optimization algorithm, which has the same optima as power EP~\citep{Minka:2004}.
This method, however, introduces some additional computational complexities compared to the local updating power EP method.
In the case both methods converge, they are expected to produce similar results.
If power EP fails to converge, it is also possible to switch on the fly to the double-loop to ensure convergence as demonstrated by \citet{Jylanki+others:2011}.

The work by \citet{Xu+others:2014} and \citet{Hasenclever+others:2017} are particular implementations of the distributed inference framework discussed in this paper.
Compared to these works, we consider the method in a more general setting and introduce some further considerations.
In addition to just EP, power EP, or SNEP, the framework can be generalized to implement other message passing techniques, all sharing the same idea of using a cavity distribution to iteratively share information between the distributed sites.
We do not argue that any particular implementation of the framework would be in general better than the other. Different implementations have different properties that are suitable and desirable in different situations.
Apart from a couple of small scale simulated experiments, we do not provide exhaustive comparison of the performance of different implementations of the distributed EP framework against each other.

We consider our method in an applied context and discuss and analyze various algorithmic considerations related to it.
For example, we discuss the implementation of sample based moment estimates for EP updates. \citet{Xu+others:2014} presented an estimate for the required parameters and conjectured that these would be unbiased, but this is not generally true.
Using experiments similar to those of \citet{Xu+others:2014}, we also study how the number of partitions affects the resulting approximation.
Compared to  conventional fully factored EP, partitioning the data into bigger subsets may yield better approximations but longer computation times.
We additionally show that dividing the data into smaller pieces tends to make the posterior variance estimates worse while maintaining the accuracy of the mean estimates. This is a known property of variational inference but has not been so well understood with EP, although it has been recognized earlier in some
form by \citet{Cunningham+others:2011} and \citet{cseke2013approximate}.

We discuss the idea in a generalized message passing framework, conforming to both the top-down and bottom-up views.
In particular, we present an efficient distributed approach for
hierarchical models, which by construction partition the data into conditionally separate pieces.
By applying EP to the posterior distribution of
the shared parameters, the algorithm's convergence only needs to happen on
this parameter subset.
We implement an example algorithm using the Stan probabilistic
programming language~\citep{Stan:2017}, leveraging its sample-based
inferences for the individual partitions.
We test the implementation in two experiments, in which we inspect the behaviour of EP in the context of the generalized framework.

The remainder of the paper proceeds as follows. We review the basic EP algorithm and introduce
terminology in Section~\ref{sec:ep}. In Section~\ref{general_framework}, we discuss the use of EP as a
general message passing framework for partitioned data, and in Section~\ref{hier}, we further demonstrate its
applicability for hierarchical models. Despite being conceptually straightforward, the implementation of an EP
algorithm involves consideration of various options in carrying out the algorithm. In Section~\ref{stuff}, we
discuss such algorithmic considerations at length, also highlighting recent methodological developments
and suggesting further generalizations. Section~\ref{experiments} demonstrates the framework with two hierarchical experiments, and Section~\ref{discussion} concludes the paper with a discussion.
Further details of implementation can be found in Appendix~\ref{appendix_sec_algorithms}.

\section{Expectation propagation}
\label{sec:ep}

The distributed inference framework presented in this paper is based on the expectation propagation algorithm. In this section, we present EP along with the more generalized idea of a message passing algorithm. Later in Section~\ref{sec:ep:consider}, we present various considerations, extensions, and related methods.

\subsection{Basic algorithm}\label{sec:ep:basic}
Expectation propagation (EP) is an iterative algorithm in which a
target density $f(\theta)$ is approximated by a density $g(\theta)$ from some
specified parametric family.
First
introduced by \citet{Opper+Winther:2000} and shortly after
generalized by \citet{Minka:2001b,Minka:2001a},
EP belongs to a group of
\emph{message passing algorithms}, which infers the target density
using a collection of localized inferences~\citep{pearl1986fusion}. In the following, we introduce the general message passing framework and then specify the features of EP.

Let us first assume that the target density $f(\theta)$ has some convenient factorization up to proportion,
\[
f(\theta) \propto \prod_{k=0}^K f_k(\theta).
\]
In Bayesian inference, the target $f$ is typically the posterior density $p(\theta|y)$, where one can assign for example factor $k=0$ to the prior $p(\theta)$ and factors 1 through K as the likelihood for the data partitioned into K parts $p(y_k | \theta)$ that are independent given the model parameters.
A message passing algorithm works by iteratively approximating
$f(\theta)$ with a density $g(\theta)$ which admits the same
factorization,
\begin{equation}
\label{eq_background_global_approx}
g(\theta) \propto \prod_{k=0}^K g_k(\theta),
\end{equation}
and using some suitable initialization for all $g_k(\theta)$. The factors $f_k(\theta)$ together with the associated approximations $g_k(\theta)$ are referred to as \emph{sites}, and the approximating distribution $g(\theta)$ is referred to as the \emph{global approximation}.

At each iteration of the algorithm, and for $k=0,\ldots,K$, we take the current approximating function $g(\theta)$ and replace $g_k(\theta)$ by the corresponding factor $f_k(\theta)$ from the target distribution. Accordingly (and with slight abuse of the term ``distribution'') we define the \emph{cavity distribution},
\[
g_{-k}(\theta) \propto \frac{g(\theta)}{g_k(\theta)},
\]
and the \emph{tilted distribution},
\[
g_{\setminus k}(\theta) \propto f_{k}(\theta) g_{-k}(\theta).
\]
The algorithm proceeds by first constructing an approximation
$g^\text{new}(\theta)$ to the tilted distribution $g_{\setminus
k}(\theta)$. After this, an updated approximation to the
target density's $f_k(\theta)$ can be obtained as
$g^\text{new}_k(\theta) \propto
g^\text{new}(\theta)/g_{-k}(\theta)$. Iterating these updates
in sequence or in parallel gives the following algorithm.
\\[2mm]
\begin{minipage}{\linewidth}
\begin{framed}
General message passing algorithm:
\begin{enumerate}
\item Choose initial site approximations $g_k(\theta)$.
\item Repeat for $k \in \{0,1,\dots,K\}$ (in serial or parallel batches) until all site approximations $g_k(\theta)$ converge:
    \begin{enumerate}
    \item \label{alg:form_cavity} Compute the cavity distribution, $g_{-k}(\theta) \propto g(\theta)/g_k(\theta)$.
    \item \label{alg:tilted_approx} Update the site approximation $g_k(\theta)$ so that $g_k(\theta) g_{-k}(\theta)$ approximates $f_{k}(\theta) g_{-k}(\theta)$.
    \end{enumerate}
\end{enumerate}
\end{framed}
\end{minipage}\\[2mm]
In some sources, step~\ref{alg:tilted_approx} above is more strictly formulated as
$$
g^\text{new}_k(\theta) = \operatorname{arg\,min}_{g_k(\theta)} \, \operatorname{D}\bigl(f_{k}(\theta) g_{-k}(\theta)\big\|g_k(\theta) g_{-k}(\theta)\bigr),
$$
where $\operatorname{D}(\cdot\|\cdot)$ corresponds to some divergence measure. In our definition, the algorithm can more freely implement any approximation method, which does not necessarily minimize any divergence.

The global approximation $g(\theta)$ and the site approximations $g_k(\theta)$ are restricted to be in a selected exponential family, such as the multivariate normal. As step~\ref{alg:tilted_approx} is usually defined, the site approximation $g_k(\theta)$ is updated so that the resulting Kullback-Leibler divergence $\operatorname{KL}\bigl(f_{k}(\theta) g_{-k}(\theta)\big\|g_k(\theta) g_{-k}(\theta)\bigr)$ is minimized.

\subsection{Further considerations}\label{sec:ep:consider}

The exponential family restriction in EP makes the algorithm efficient: any product and division between these distributions stays in the parametric family and can be carried out analytically by summing and subtracting the respective natural parameters.
The complexity of these distributions, which is determined by the number of parameters in the model, remains constant regardless of the number of sites.
This is less expensive than carrying around the full likelihood, which in general would require computation time proportional to the size of the data.
Accordingly, EP tends to be applied to specific high-dimensional problems where computational cost is an issue, notably for Gaussian processes~\citep{Rasmussen+Williams:2006,Jylanki+others:2011,Cunningham+others:2011,Vanhatalo+others:2013,cseke2013approximate}, and efforts are made to keep the algorithm both stable and fast.

Approximating the tilted distribution in step~\ref{alg:tilted_approx} is, in many ways, the core step of a message passing algorithm. In EP, this is typically done by matching the moments of $g_k(\theta) g_{-k}(\theta)$ to those of $f_{k}(\theta) g_{-k}(\theta)$, which corresponds to minimizing the Kullback-Leibler divergence
$\operatorname{KL}(g_{\backslash k}(\theta)||g(\theta))$.
In Section \ref{approximation}, we discuss in more detail a variety of other choices for forming tilted approximations, beyond the standard choices in the EP literature.
If $f_{k}(\theta)$ has the same form as $g$ then the contribution of that term can be computed exactly and there is no need for the corresponding site approximation term $g_{k}(\theta)$. For example, if the prior $f_0(\theta)$ and approximating distribution $g$ are both multivariate normals, then only tilted distributions $k=1,\ldots,K$ need to be computed.

Even if EP minimizes local KL-divergence in the scope of each site, it will not in general minimize the KL-divergence from the target density to the global approximation $\operatorname{KL}(f(\theta)||g(\theta))$.
Furthermore, there is no general guarantee of convergence for EP. However, for models with log-concave factors $f_k$ and initialization to the prior distribution, the algorithm has proven successful in many applications.
Various studies have been made to assess the behaviour of EP.
\citet{dehaene2015bounding} present bounds for the approximate error. \citet{dehaene2018expectation} inspect the method in the large data limit and show that it is asymptotically exact but it may diverge if initialized poorly. \citet{dehaene2016expectation} relate EP to other better understood methods and show that it is equivalent to performing gradient descent on a smoothed energy landscape.

Generally, message passing algorithms require that the site distributions $g_k(\theta)$ are stored in memory, which may be a problem with a large number of sites.
\citet{dehaene2018expectation} and \cite{li:2015} present a modified EP method in which sites share the same approximate factor $g_{\,\text{site}}(\theta)$;
considering the prior $p(\theta)$ as a constant site with index 0, and setting all the other site approximations $g_k(\theta) = g_{\,\text{site}}(\theta), k=1,2,\dots,K$, the global approximation becomes $g(\theta) \propto p(\theta) g_{\,\text{site}}(\theta)^K$.
While making the algorithm more memory efficient, it has been shown for certain applications that the method works almost as well as the original EP.

\section{Message passing framework for partitioned data}
\label{general_framework}

\begin{figure}
\centerline{\includegraphics{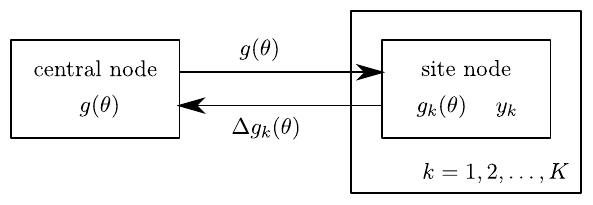}}
\caption{%
The EP framework for partitioned data.
The central node stores the current parameters for the global approximation $g(\theta)$.
Each site node $k=1,2,\dots,K$ stores the current parameters for the site approximation $g_k(\theta)$ and the assigned partition of the data $y_k$.
The central node sends the parameters of $g(\theta)$ to the site nodes.
In parallel, the site nodes update $g_k(\theta)$ and send back the difference in the parameters.
}\label{fig_partitioned_schema}
\end{figure}

The factorized nature of the EP algorithm defined in Section~\ref{sec:ep} makes it a suitable tool for partitioned data.
Assuming the likelihood factorizes over the partitions, the likelihood of each part can be assigned to its own site:
\begin{align}
    p(\theta | y) \propto p(\theta) \prod_{k=1}^K p(y_k | \theta),
\end{align}
where each term $p(y_k | \theta)$ with respective data partition $y_k=[y_{k,1}, y_{k,2}, \dots,y_{k,n_k}]$ is approximated with site approximation $g_k(\theta)$.
The algorithm can be run in a distributed setting consisting of a central node and site nodes.
The schema is illustrated in Figure~\ref{fig_partitioned_schema}.
The central node stores the current global approximation $g(\theta)$ and controls the messaging for the sites, while each site node stores the corresponding partition of the data $y_k$ and the current site approximation $g_k(\theta)$.
The central node initiates the updates by sending the natural parameters of the current global approximation $g(\theta)$ to the sites.
Given this information, the sites update the respective site approximations $g_k(\theta)$ and send back the change in the natural parameters of the site distribution:
\begin{align*}
    g_{-k}(\theta) &\propto g(\theta) / g_k(\theta)
    && \text{subtraction of natural parameters} \\
    g_{\setminus k}(\theta) &\propto p(y_k|\theta) g_{-k}(\theta)
    && \text{MCMC sampling} \\
    g^\text{new}(\theta) &\approx g_{\setminus k}(\theta)
    && \text{parameter sample estimates} \\
    \mathrm{\Delta} g_k(\theta) &\propto g^\text{new}(\theta)/g(\theta)
    && \text{subtraction of natural parameters}.
\intertext{The central node then receives the differences and aggregates these to update the global approximation by adding in the received parameter changes:
}
    g^\text{new}(\theta) &\propto g(\theta) \mathrm{\Delta} g_k(\theta)
    && \text{sum of natural parameters}.
\end{align*}
This enables model parallelism---in that each site node can work
independently to infer its assigned part of the model---and data
parallelism---in that each site node only needs to store its assigned
data partition~\citep{dean2012large}.  We present the algorithm in more detail in Appendix~\ref{appendix_parallelep}.

In a conventional EP setting, the likelihood is factorized pointwise so that each site corresponds to one data point. This is motivated by the simplicity of the resulting site updates, which can often be carried out analytically. By assigning multiple data points to one site, the updates become more difficult and time consuming. However, updating such a site also provides more information to the global approximation and the algorithm may converge in fewer iterations. In addition, the resulting approximation error should be smaller as the number of sites decreases.

In EP, as mentioned in Section~\ref{sec:ep}, approximating the tilted distribution in step~\ref{alg:tilted_approx} of the general message passing algorithm is carried out by moment matching. This makes EP particularly useful in the context of partitioned data: intractable site updates can be conveniently inferred by estimating the tilted distribution moments, for example using MCMC. Other message passing algorithms, where some other method for tilted distribution approximation is used, can also be applied in such a context. These are discussed in more detail in Section~\ref{approximation}.

\begin{figure}[p]
\centerline{\includegraphics[width=.35\textwidth]{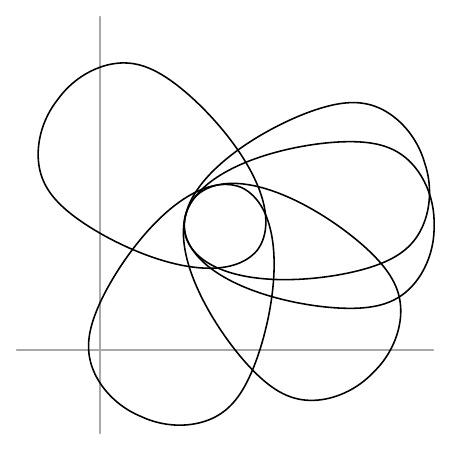}}
\caption{Sketch illustrating the benefits of message passing in Bayesian computation.  In this simple example, the parameter space $\theta$ has two dimensions, and the data have been split into five pieces.  Each oval represents a contour of the likelihood $p(y_k | \theta)$ provided by a single partition of the data.  A simple parallel computation of each piece separately would be inefficient because it would require the inference for each partition to cover its entire oval.  By combining with the cavity distribution $g_{-k}(\theta)$, we can devote most of our computational effort to the area of overlap.}\label{sketch}
\end{figure}

\begin{figure}[p]
\centerline{\includegraphics[width=.6\textwidth]{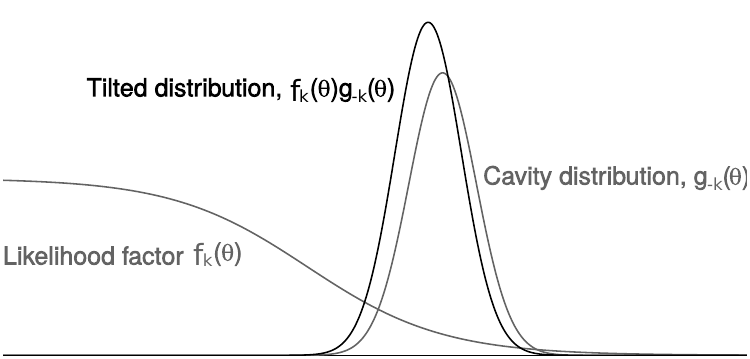}}
\caption{Example of a step of an EP algorithm in a simple one-dimensional example, illustrating the stability of the computation even when part of the likelihood is far from Gaussian.  When performing inference on the likelihood factor $p(y_k|\theta)$, the algorithm uses the cavity distribution $g_{-k}(\theta)$ as a prior.}\label{separation}
\end{figure}

In divide-and-conquer algorithms, each partition of the data is processed separately and the results are combined together in a single pass. This behavior resembles the first iteration of the EP algorithm. In EP however, the global approximation is further optimized by iteratively updating the sites with shared information from the other sites. In contrast to divide-and-conquer algorithms, each step of an EP algorithm combines the likelihood of one partition with the cavity distribution representing the rest of the available information across the other $K-1$ pieces (and the prior). This extra information can be used to concentrate the computational power economically in the areas of interest. Figure \ref{sketch} illustrates this advantage with a conceptual example, showing how the inference for each site factor $f_k(\theta)$ can be concentrated in a region where all site factors overlap.
Figure \ref{separation} illustrates the construction of the tilted distribution $g_{\backslash k}(\theta)$ and demonstrates the critically important regularization attained by using the cavity distribution $g_{-k}(\theta)$ as a prior; because the cavity distribution carries information about the posterior inference from all other $K-1$ data pieces, any computation done to approximate the tilted distribution (step~\ref{alg:tilted_approx} in the message passing algorithm) will focus on areas of greater posterior mass.

\section{Application to hierarchical models}\label{hier}

In a hierarchical context, EP can be used to efficiently divide a multiparameter problem into sub-problems with fewer parameters. If the data assigned to one site are not affected by some parameter, the site does not need to take this local parameter into account in the update process. By distributing hierarchical groups into separate sites, the sites can ignore the local parameters from the other groups.

\subsection{Posterior inference for the shared parameters}\label{subsec:shared_param_inf}

\begin{figure}
\centering \includegraphics{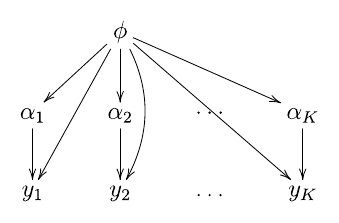}
\caption{ Model structure for the hierarchical EP algorithm. In each site $k$, inference is based on the local model, $p(y_k|\alpha_k,\phi)p(\alpha_k|\phi)$, multiplied by the cavity distribution $g_{-k}(\phi)$. Computation on this tilted posterior gives a distributional approximation on $(\alpha_k,\phi)$ or simulation draws of $(\alpha_k,\phi)$; in either case, we just use the inference for $\phi$ to update the local approximation $g_k(\phi)$. The algorithm has potentially large efficiency gains because, in each of the $K$ sites, both the sample size and the number of parameters scale proportional to $1/K$.}\label{tree}
\end{figure}

Suppose a hierarchical model has local parameters $\alpha_1,\alpha_2,\dots,\alpha_K$ and shared parameters $\phi$. All these can be vectors, with each $\alpha_k$ applying to the model for the data piece $y_k$, and with $\phi$ including shared parameters of the data model and hyperparameters as well.
This structure is displayed in Figure \ref{tree}. Each data piece $y_k$ is assigned to one site with its own local model $p(y_k|\alpha_k,\phi)p(\alpha_k|\phi)$. The posterior distribution is
\begin{align}
\label{eq_hier_true_joint_posterior}
\begin{split}
p(\phi, \alpha | y) &\propto p(\phi, \alpha) p(y|\phi, \alpha)
= p(\phi) p(\alpha|\phi) p(y|\phi, \alpha) \\
&= p(\phi) \prod_{k=1}^K p(y_k|\alpha_k, \phi) p(\alpha_k | \phi),
\end{split}
\end{align}
where $\alpha = (\alpha_1,\alpha_2,\dots,\alpha_K)$.

As each local parameter $\alpha_k$ affects only one site, they do not need to be included in the propagated messages.  EP can thus be applied to approximate the marginal posterior distribution of $\phi$ only. If desired, the joint posterior distribution of all the parameters can be approximated from the obtained marginal approximation with the methods discussed later in Section~\ref{subsec:joint_approx}.

Applying EP for the marginal posterior distribution $p(\phi|y)$ is straightforward. Marginalizing the joint posterior distribution in~\eqref{eq_hier_true_joint_posterior} gives
$$
p(\phi|y) = \int p(\phi, \alpha | y) \:\mathrm{d}\alpha
\propto p(\phi) \prod_{k=1}^K \int p(y_k|\alpha_k,\phi)p(\alpha_k|\phi) \:\mathrm{d}\alpha_k,
$$
which is approximated by
$$
p(\phi|y) \approx g(\phi) = p(\phi) \prod_{k=1}^K g_k(\phi).
$$
Given the cavity distribution $g_{-k}(\phi)$, each site $k$ approximates the tilted distribution
\begin{align}
\label{eq_hier_tilted_distribution}
g_{\setminus k}(\phi)
\propto \int g_{-k}(\phi) p(y_k|\alpha_k,\phi)p(\alpha_k|\phi) \:\mathrm{d}\alpha_k
\end{align}
in the restricted exponential family form by determining its moments, after which the site updates the respective approximation $g_k(\phi)$ accordingly. For intractable tilted distributions, as is often the case, simulation-based methods provide a practical general approach.

The computational advantage of this marginalized approach is that the local parameters $\alpha$ are partitioned. For example, suppose we have a model with 100 data points in each of $3\,000$ groups, 2 local parameters per group (a varying slope and intercept) and, say, 20 shared parameters (including fixed effects and hyperparameters). If we divide the problem into $K=300$ sites with 10 groups each, we have reduced a problem with $300\,000$ data points and $6\,020$ parameters to $300$ parallel iterated problems with $1000$ data points and $40$ parameters (20 local and 20 shared parameters) each.

\subsection{Posterior inference for the other parameters}\label{subsec:joint_approx}
In large-dimensional hierarchical scenarios, the full joint posterior distribution is not typically needed. If all that is required are the marginal posterior distributions for each $\alpha_k$ separately, we can take these directly from the corresponding tilted distribution inferences from the last iteration. The marginal posterior distribution for local parameter $\alpha_k$ can be obtained from the joint distribution in~\eqref{eq_hier_true_joint_posterior} by
\begin{align*}
p(\alpha_k|y)
&= \int \int_{\alpha \setminus \alpha_k} p(\phi, \alpha | y) \:\mathrm{d}\alpha_p \:\mathrm{d}\phi \\
&\propto \int p(\phi) p(y_k|\alpha_k,\phi)p(\alpha_k|\phi) \prod_{p\neq k} \int p(y_p|\alpha_p,\phi)p(\alpha_p|\phi) \:\mathrm{d}\alpha_p \:\mathrm{d}\phi.
\intertext{Assuming the EP algorithm has converged, this can be approximated:}
p(\alpha_k|y) &\approx \int g_{-k}(\phi) p(y_k|\alpha_k,\phi)p(\alpha_k|\phi) \:\mathrm{d}\phi,
\end{align*}
which is the same as the tilted distribution in~\eqref{eq_hier_tilted_distribution} but marginalized over $\phi$ instead of $\alpha_k$. If, for example, a sample-based method is used for the tilted distribution inference in EP, one can easily just store the local parameter samples in the last iteration to form the marginal posterior distribution for them.

If the joint posterior distribution of all the parameters is required, one can approximate it using the obtained EP approximation $g(\phi)$ for the marginal posterior distribution of the shared parameters:
$$
p(\phi, \alpha | y)
= p(\phi | y) p(\alpha | \phi, y)
= p(\phi | y) \prod_{k=1}^K p(\alpha_k | \phi, y_k)
\approx g(\phi) \prod_{k=1}^K p(\alpha_k | \phi, y_k).
$$
To get simulation draws from this, one can first take some number of draws from $g(\phi)$, and then, for each draw, run $K$ parallel MCMC inferences for each $\alpha_k$ conditional on the sampled value of $\phi$.
This computation is potentially expensive---for example, to perform it using 100 random draws of $\phi$ would require 100 separate MCMC inferences---but, on the plus side, each run should converge fast because it is conditional on the hyperparameters of the model.
In addition, it may ultimately be possible to use adiabatic Monte Carlo~\citep{Betancourt:2014} to perform this ensemble of simulations more efficiently.

\section{Algorithmic considerations}\label{stuff}

This section discusses various details related to the implementation of an EP or message passing algorithm in general. Some of the key aspects to consider are:
\begin{itemize}
\item \textbf{Partitioning the data.}
From the bottom-up view, such as with private data, the number of partitions $K$ is simply given by the number of data owners. From the top-down view with distributed computing, $K$ will be driven by computational considerations.  If $K$ is too high, the site approximations may not be accurate.  But if $K$ is low, then the computational gains will be small.  For large problems it could make sense to choose $K$ iteratively, for example starting at a high value and then decreasing it if the approximation seems too poor.  Due to the structure of modern computer memory, the computation using small blocks may get additional speed-up if most of the memory accesses can be made using fast but small cache memory.

\item \textbf{Parametric form of the approximating distributions
$g_k(\theta)$.}
The standard choice is the multivariate normal family, which will also work
for any constrained space with appropriate transformations; for example, one can use logarithm for all-positive and logit for interval-constrained parameters.
For simplicity we may also assume that the prior distribution $p_0(\theta)$ is multivariate normal, as is the case in many practical applications, sometimes after proper reparameterization. Otherwise, one may treat the prior as an extra site which will also be iteratively approximated by some Gaussian density $g_0$. In that case, some extra care is required regarding the initialization of $g_0$.
We will discuss alternative options in Section \ref{sub:families}.

\item \textbf{Initial site approximations $g_k$.}
One common choice is to use improper uniform distributions. With normal approximation, this corresponds to setting natural parameters to zeros. Alternatively, one could use a broad but proper distribution factored into $K$ equal parts, for example setting each $g_k(\theta)=\mbox{N}(0,\frac{1}{K}A^2I)$, where $A$ is some large value (for example, if the elements of $\theta$ are roughly scaled to be of order 1, we might set $A=10$).

\item \textbf{Algorithm to perform inference on the tilted distribution.}
We will discuss three options in Section \ref{approximation}:
deterministic mode-based approximations, divergence measure
minimizations, and Monte Carlo simulations.

\item \textbf{Asynchronous site updates.}
In a distributed context, particularly with unevenly sized data partitions, it can be beneficial to allow a site to be updated as soon as it has finished its previous update, even if some other sites are still busy. Different rules for waiting for more information could be applied here, as long as it is ensured that at least one other site is updated before starting the next iteration.

\item \textbf{Improper site distributions.}
When updating a site term $g_k$ in step~\ref{alg:tilted_approx} in the message passing algorithm, the division by the cavity distribution can yield a covariance or precision matrix that is not positive definite. This is not a problem in itself as the site approximations do not need to be proper distributions. However, improper site distributions may lead to improper global approximations or tilted distributions in subsequent iterations, which is a problem. Various methods for dealing with this issue are discussed in Section \ref{eigens}.

\end{itemize}
In the following sections, we address some of these issues in detail,
namely, how to approximate the tilted distribution and how to handle
potential numerical instabilities in the algorithms. The methods and aspects discussed in this section cover multiple different implementations for the distributed EP method. Different methods may work in different situations and, as in statistical analysis in general, one has to choose one that suits the problem.
In this paper, we present all the prominent approaches in a high level while focusing in one implementation, where the tilted distribution inference is carried out by sampling.
With this approach, the inference can be carried out conveniently with probabilistic programming tools, which provides substantial generalizability.

\subsection{Approximating the tilted distribution}\label{approximation}

In EP, the tilted distribution approximation in step~\ref{alg:tilted_approx} is
framed as a moment matching problem, where attention is restricted to
approximating families estimable with a finite number of moments.
For example, with the multivariate normal family, one chooses the site
$g_k(\theta)$ so that the first and second moments of
$g_k(\theta)g_{-k}(\theta)$ match those of the possibly intractable tilted
distribution $g_{\backslash k}(\theta)$. When applied to Gaussian
processes, this approach has the particular advantage that the tilted
distribution $g_{\backslash k}(\theta)$ can typically be set up as a
univariate distribution over only a single dimension in $\theta$.
This dimension reduction implies that the tilted distribution
approximation can be performed analytically
\citep[e.g.][]{Opper+Winther:2000,Minka:2001b}
or relatively quickly using one-dimensional
quadrature~\citep[e.g.][]{Zoeter+Heskes:2005}.
In higher dimensions, quadrature gets computationally more expensive or, with a reduced number
of evaluation points, the accuracy of the moment computations gets
worse. \citet{Seeger+Jordan:2004} estimated the tilted moments in multiclass
classification using multidimensional quadratures.  Without the
possibility of dimension reduction in the more general case,
approximating the integrals to obtain the required moments over $\theta \in \reals^k$ becomes a hard task.

To move towards a black-box message passing algorithm, we inspect the tilted distribution approximation from
four perspectives:
matching the mode, minimizing a divergence measure, using numerical simulations, and using nested EP.
Algorithmically, these correspond to Laplace methods, variational
inference, Monte Carlo, and recursive message passing, respectively.
Critically, the resulting algorithms preserve the essential idea that the local pieces of data are analyzed
at each step in the context of a full posterior approximation.

\subsubsection*{Mode-based tilted approximations}\label{mode}
The simplest message passing algorithms construct an approximation of the
tilted distribution around its mode at each step. As Figure \ref{separation} illustrates, the tilted distribution can have a well-identified mode even if the factor of the likelihood does not.

An example of a mode-based approximation is obtained by, at each step, setting
$g^{\rm new}$ to be the (multivariate) normal distribution centered at
the mode of $g_{\backslash k}(\theta)$, with covariance matrix equal
to the inverse of the negative Hessian of $\log g_{\backslash k}$ at
the mode.
This corresponds to the Laplace approximation, and the
message passing algorithm corresponds to Laplace propagation~\citep{Smola+others:2004}.
The proof presented by \citet{Smola+others:2004} suggests that a fixed point of Laplace propagation corresponds to a local mode of the joint model and hence also one possible Laplace approximation. Therefore, with Laplace approximation, a message passing algorithm based on local approximations corresponds to the global solution. \citet{Smola+others:2004} were able to get useful results with tilted distributions in several hundred dimensions.
The method has been shown to work well in many problems~\citep[see e.g.][]{Rue+others:2009}.

The presence of the cavity distribution as a prior  (as illustrated in Figure \ref{separation}) gives two computational advantages to this algorithm.  First, we can use the prior mean as a starting point for the algorithm; second, the use of the prior ensures that at least one mode of the tilted distribution will exist.

To improve upon this simple normal approximation, we can evaluate the tilted distribution at a finite number of points around the mode and use this to construct a better approximation to capture asymmetry and long tails in the posterior distribution.  Possible approximate families include the multivariate split-normal~\citep{Geweke:1989,Villani+Larsson:2006}, split-$t$, or wedge-gamma~\citep{Gelman+others:2014} distributions.
We are {\em not} talking about changing the family of approximate distributions $g$---we would still keep these as multivariate normal.  Rather, we would use an adaptively-constructed parametric approximation, possibly further improved by importance sampling~\citep{Geweke:1989,Vehtari+Gelman+Gabry:2016-PSIS} or central composite design integration~\citep{Rue+others:2009} to get a better approximation of the moments of the tilted distribution to used in constructing $g_k$.

\subsubsection*{Variational tilted approximations}
\label{considerations:variational_tilted_approx}
Mode-finding message passing algorithms have the advantage of simplicity, but they can do a poor job at capturing uncertainty when approximating the tilted distribution. An alternative
approach is to find the closest
distribution within an approximating family to the tilted distribution, using
a divergence measure to define closeness.
If the approximating family
contains the tilted distribution as one member in the family, then
the local inference is exact (step~\ref{alg:tilted_approx} in the algorithm).
In practice, this is not the case, and the behavior of the local
variational approximations depends on the properties of the
chosen divergence measure.
This generalizes mode-finding, which corresponds to minimizing a particular divergence measure.

In the classical setup of EP, the chosen divergence measure is the Kullback-Leibler divergence from the tilted distribution to the global approximation,
$\operatorname{KL}(g_{\backslash k}(\theta)||g^{\rm new}(\theta))$.
As discussed before in Section~\ref{sec:ep}, if the approximating distribution
forms an exponential family, minimizing the divergence conveniently corresponds to matching the moments of two distributions~\citep{Minka:2001b}.

Another reasonable divergence measure is the reverse KL divergence from the global approximation to the tilted distribution,
$\operatorname{KL}(g^{\rm new}(\theta)||g_{\backslash k}(\theta))$.
This is known as variational message passing~\citep{Winn+Bishop:2005}, where the local computations to
approximate the tilted distribution can be shown to maximize a
lower bound on the marginal likelihood. In fact, variational message passing enjoys
the property that the algorithm minimizes
a global divergence to the posterior, $\operatorname{KL}(g(\theta)||p(\theta|y))$,
according to the factorized approximating family
$g(\theta)=p(\theta)\prod_{k=1}^Kg_k(\theta)$.

Inference can also be done using the $\alpha$-divergence family,
in which $\alpha=1$ corresponds to the KL divergence used in the
classical EP, $\alpha=0$ corresponds to the reverse KL divergence, and
$\alpha=0.5$ corresponds to Hellinger distance.
One algorithm to solve this is known as power EP~\citep{Minka:2004}.
Power EP has been shown to improve the robustness of the algorithm when the approximation family is not flexible enough~\citep{Minka:2005} or when the propagation of information is difficult due to vague prior information~\citep{Seeger:2008}.
This can be useful when moment computations are not accurate, as
classical EP may have stability issues~\citep{Jylanki+others:2011}. Even with one-dimensional tilted distributions, moment computations are more challenging if the tilted distribution is multimodal or has long tails.
Ideas of power EP in general might help to stabilize message passing
algorithms that use approximate moments, as $\alpha$-divergence with $\alpha<1$ is less sensitive to errors in tails of the approximation.

\subsubsection*{Energy optimization}\label{energy_optimization}

The EP algorithm, like message passing algorithms in general, is not guaranteed to converge.
It is possible, however, to define an objective function whose stationary points corresponds to a fixed point for the EP algorithm.
In its general form, the problem can be formulated as an optimization for the free energy corresponding to the negative logarithm of the intractable normalizer of the global approximation in Equation~\eqref{eq_background_global_approx}~\citep{Opper+Winther:2005}.
Appendix~\ref{appendix_ep_energy} illustrates such a formulation in more detail.
\citet{heskes2005} presents an unifying analysis of the correspondence of various different formulations of the same objective, and \citet{dehaene2016expectation} relates EP to using a gradient descent on a smoothed energy landscape.
Various energy optimization methods, for which convergence is guaranteed, can be applied to directly find analogous approximations.
\citet{Heskes+Zoeter:2002} show a simulated example where EP fails to converge but a double-loop optimization algorithm is successful.
While optimising similar objective functions and possibly admitting similar distributed local updating frameworks, these algorithms are often slower.

Based on the original EP min-max optimization problem reviewed in Appendix~\ref{appendix_ep_energy}, \citet{Opper+Winther:2005} derived a convergent double-loop optimization algorithm called expectation consistent approximate inference (EC).
Recently \citet{Hasenclever+others:2017} presented a similar but faster double-loop optimization algorithm called stochastic natural gradient expectation propagation (SNEP), that shares the same optimum as power EP, and which admits a similar distributed framework as the local updating scheme discussed in this paper.
They show that, instead of the natural parameter space, SNEP can be seen as a mean parameter space version of the damped EP update.
In the case of convergence, both methods are expected to produce similar results.
We briefly compare the methods in a simulated experiment in Appendix~\ref{appendix_snep_vs_ep} with simulation based site inferences. These experiments show that moment matching can reach convergence faster but may suffer from larger variability. SNEP can be slower and behave chaotically when far from convergence but tends to have smaller variability when reaching stable progression and eventually convergence.
It is also possible to apply moment matching in early iterations in order to have good initial progression and switch to SNEP for more stable convergence after some iterations. \citet{Jylanki+others:2011} present similar ideas in a conventional EP setting in order to ensure convergence.
However, convergence problems might also indicate that the approximating family matches poorly with the exact posterior~\citep{Minka:2001b,Jylanki+others:2011}.
Thus it could be beneficial to first consider some alternative approximating families.

Black-box $\alpha$-divergence minimization~\citep{Hernandez:2016} (BB-$\alpha$) is another example of an energy optimizing algorithm with a tunable $\alpha$-divergence measure and automatic differentiation. Similar to the stochastic EP method by~\citet{li:2015} discussed in Section~\ref{sec:ep:consider}, the BB-$\alpha$ method features shared local approximations, which makes it more memory efficient. In distributed settings, however, each site unit needs to store a copy of the shared site distribution anyway. Thus no memory is saved by tying up the factors in this setting.

\subsubsection*{Simulation-based tilted approximations}\label{hmc}
An alternative approach is to
use simulations (for example, Hamiltonian Monte Carlo using Stan) to
approximate the tilted distribution at each step and then use these to set the
moments of the approximating family. As above, the advantage of the EP message passing algorithm here is that the computation only uses a fraction $1/K$ of the data, along with a simple exponential family prior (typically multivariate normal on parameters that if necessary have been transformed to an unbounded scale) that comes from the cavity distribution.

As with methods such as stochastic variational inference~\citep{Hoffman+others:2013} which take steps based on stochastic estimates, the properties of the estimator affect the convergence properties of the EP algorithm.
One way to study convergence is to inspect the expectation of the state of the algorithm at the fixed point of conventional analytic EP.
As discussed in Section~\ref{sec:ep}, in the EP update step the KL divergence from the new global approximation to the tilted distribution $\operatorname{KL}(g_{\backslash k}(\theta)||g(\theta))$ is minimized by matching the moments.
With a simulation-based method, the expectation of the new global approximation moments in step~\ref{alg:tilted_approx} should then match with the tilted distribution moments.
When working with the normal approximation, we would use the unbiased estimates of the mean and covariance of the tilted distribution, which are easily obtained from the simulated sample.
Using this estimator would not result in the least possible expected KL divergence in general, however.
In addition, in the algorithm, these parameters are ultimately needed in natural form, and estimating them is a complex task in general.
This problem is discussed in more detail in Appendix~\ref{app_precision}.
If needed, the variance of the estimates can be reduced while preserving unbiasedness by using control variates.
While MCMC computation of the moments may give inaccurate estimates, we suspect that they will work better than, or as a supplement to, a Laplace approximation for skewed distributions.
%TODO: REF

With sample based estimates, there is a tradeoff between computation time and precision of the estimates.
In the local bottom-up view of a distributed inference problem, the time taken for the separate inferences is not crucial. Thus it is appropriate to apply MCMC with suitably large sample sizes for such problems.

In serial or parallel EP, samples from previous iterations can be reused as starting points
for Markov chains or in importance sampling. We discuss briefly the latter.
Assume we have obtained at iteration $t$ for node $k$, a set of posterior simulation draws
$\theta_{t,k}^s$, $s=1,\ldots,S_{t,k}$
from the tilted distribution $g_{\backslash k}^t$, possibly with
weights $w_{t,k}^s$; take  $w_{t,k}^s\equiv 1$ for an unweighted sample.
To progress to node $k+1$, reweight
these simulations as:
$ w_{t,k+1}^s = w_{t,k}^s  g_{\backslash (k+1)}^t(\theta_{t,k}^s)/g_{\backslash k}(\theta_{t,k}^s)$.
If the vector of new weights has an effective sample size,
$$ \mathrm{ESS} = \frac{\left(\frac{1}{S}\sum_{s=1}^S w^s_{t,k+1}\right)^2}{\frac{1}{S}\sum_{s=1}^S (w^s_{t,k+1})^2}, $$
that is large enough, keep this sample, $\theta_{t,k+1}^s=\theta_{t,k}^s$. Otherwise
throw it away, simulate new  $\theta_{t+1,k}^s$'s from $g_{\backslash k+1}^t$,
and reset the weights $w_{t,k+1}$ to 1.
This basic approach was used in the EP-ABC algorithm of~\citet{Barthelme+Chopin:2014}.
Furthermore, instead of throwing away a sample with too low ESS, one could move these through several MCMC steps, in the spirit of sequential Monte Carlo~\citep{DelMoral+others:2006}.
Another approach, which can be used in serial or parallel EP, is to use adaptive multiple importance sampling~\citep{Cornuet+others:2012}, which would make it possible to recycle the simulations from previous iterations.
Even the basic strategy should provide important savings when
EP is close to convergence. Then changes in the tilted distribution should become small and as a result the variance of the importance weights should be small as well.
In practice, this means that the last EP iterations should essentially come for free.

\subsubsection*{Nested EP}
In a hierarchical setting, the model can be fit using the nested EP approach~\citep{Riihimaki+others:2013, hernandez_hernandez_2016}, where moments of the tilted distribution are also estimated using EP.
This approach leads to recursive message passing algorithms, often applied in the context of graphical models, where the marginal distributions of all the model parameters are inferred by passing messages along the edges of the graph~\citep{Minka:2005} in a distributed manner.
As in the hierarchical case discussed in Section~\ref{hier}, the marginal approximation for the parameters can be estimated without forming the potentially high-dimensional joint approximation of all unknowns. This framework can be combined together with other message passing methods, adopting suitable techniques for different parts of the model graph. This distributed and extendable approach makes it possible to apply message passing to arbitrarily large models~\citep{wand2017fast}.

\subsection{Damping}\label{sec_damping}

As mentioned in Section~\ref{sec:ep}, although the EP algorithm iteratively minimizes the KL divergences from the tilted distributions to their corresponding approximations, it does not minimize the KL divergence from the target density to the global approximation.
In particular, running the EP updates in parallel often yields a deviated global approximation when compared to the result obtained with sequential updates~\citep{Minka+Lafferty:2002,Jylanki+others:2011}.
In order to fix this problem, damping can be applied to the site approximation updates.

Damping is a simple way of performing an EP update on the site distribution only partially by reducing the step size. Consider a damping factor $\delta \in (0,1]$. A partially damped update can be carried out by,
$$
g^\text{new}_k(\theta) = g_k(\theta)^{1-\delta} \left( \widetilde{g}_{\setminus k}(\theta) / g_{-k}(\theta) \right)^\delta,
$$
where $\widetilde{g}_{\setminus k}(\theta)$ is the corresponding tilted distribution approximation. This corresponds to scaling the difference in the natural parameters of $g_k(\theta)$ by $\delta$. When $\delta=1$, no damping is applied at all.

The error in the parallel EP approximation can be avoided by using a small enough damping factor $\delta$. However, this reduction in the step size makes the convergence slower and thus it is beneficial to keep it as close to one as possible. The amount of damping needed varies from problem to problem and it can often be determined by testing. \citet{Minka+Lafferty:2002} proposes to set $\delta = 1/K$ as a safe rule.
However, with a large number of sites $K$, this often results in intolerably slow convergence.
In order to speed up the convergence, it could be possible to start off with damping closer to 1 and decrease it gradually with the iterations without affecting the resulting approximation.
In our experiments, by comparing the resulting approximation to a known target, we found out that in the first iteration, $\delta = 0.5$ often resulted in good progression, regardless of the number of sites $K$. In the following iterations, we obtained good results by decreasing damping gradually to $\delta = 1/K$ in $K$ iterations.

In addition to fixing the approximation error, damping helps in dealing with some convergence issues, such as oscillation and non-positive-definiteness in approximated parameters. If these problems arise with the selected damping level, one can temporarily decrease it until the problem is solved, and this step can be automated.

\subsection{Keeping the covariance matrix positive definite}\label{eigens}
In EP, it is not required that the site approximations be proper distributions. They are approximating a likelihood factor, not a probability distribution, at each site.
Tilted distributions and the global approximation, however, must be proper, and situations where these would become improper must be addressed somehow. These problems can be caused by numerical instabilities and also can also be inherent to the algorithm itself.

As discussed before, obtaining the updated site distribution from an approximated tilted distribution in step~\ref{alg:tilted_approx} of the message passing algorithm, can be conveniently written in terms of the natural parameters of the exponential family:
\begin{align*}
Q^{\rm new}_k = Q^{\rm new}_{\setminus k} - Q_{-k},  \qquad
r^{\rm new}_k = r^{\rm new}_{\setminus k} - r_{-k}, \nonumber
\end{align*}
where each $Q=\Sigma^{-1}$ denote the precision matrix and each $r=\Sigma^{-1}\mu$ denote the precision mean of the respective distribution.
Here the approximated natural parameters $Q^{\rm new}_{\setminus k}$ and $r^{\rm new}_{\setminus k}$ of the tilted distribution together with the parameters $Q^{\rm new}_{- k}$ and $r^{\rm new}_{- k}$ of the cavity distribution are being used to determine the new site approximation parameters $Q^{\rm new}_{k}$ and $r^{\rm new}_{k}$.
As the difference between the two positive definite matrices is not itself necessarily positive definite, it can be seen that the site approximation can indeed become improper.

Problems with the tilted distribution can arise when many of the site approximations become improper. Constraining the sites to proper distributions (perhaps with the exception of the initial site approximations) can fix some of these problems~\citep{Minka:2001b}. In the case of a multivariate normal distribution, this corresponds to forcing the covariance or precision matrix to be positive definite. If all the sites are positive definite, all the cavity distributions and the global approximation will also be positive definite.

The simplest way of dealing with non-positive definite matrices is to simply ignore any update that would lead into such and hope that future iterations will fix this issue. Another simple option is to set the covariance matrix $\Sigma^{\rm new}_k = a I$ with some relatively big $a$ and preserve the mean.

Various methods exist for transforming a matrix to become positive definite. One idea, as in the SoftAbs map of~\citet{Betancourt:2013}, is to do an eigendecomposition, keep the eigenvectors but replace all negative eigenvalues with a small positive number and reconstruct the matrix. Another possibly more efficient method is to find only the smallest eigenvalue of the matrix and add its absolute value and a small constant to all the diagonal elements in the original matrix. The former method is more conservative, as it keeps all the eigenvectors and positive eigenvalues intact, but it is computationally heavy and may introduce numerical error. The latter preserves the eigenvectors but changes all of the eigenvalues. However, it is computationally more efficient. If the matrix only slightly deviates from positive definite, it is justified to use the latter approach as the change on the eigenvalues is not big. If the matrix has big negative eigenvalues, it is probably best not to try to modify it in the first place.

If damping is used together with positive definite constrained sites, it is only necessary to constrain the damped site precision matrix, not the undamped one. Because of this, it is possible to find a suitable damping factor $\delta$ so that the update keeps the site, or all the sites in parallel EP, positive definite. This can also be used together with other methods, for example by first using damping to ensure that most of the sites remain valid and then modifying the few violating ones.

\subsection{Different families of approximate distributions}
\label{sub:families}

 We can place the EP approximation, the tilted distributions, and the target distribution on different rungs of a ladder:
 \begin{itemize}
 \item $g=p_0\prod_{k=1}^K g_k$, the EP approximation;
 \item For any $k$, $g_{\backslash k}=g\frac{p_k}{g_k}$, the tilted distribution;
 \item For any $k_1,k_2$, $g_{\backslash k_1,k_2}=g\frac{p_{k_1}p_{k_2}}{g_{k_1}g_{k_2}}$, which we might call the tilted$^2$ distribution;
 \item For any $k_1,k_2,k_3$, $g_{\backslash k_1,k_2,k_3}=g\frac{p_{k_1}p_{k_2}p_{k_3}}{g_{k_1}g_{k_2}g_{k_3}}$, the tilted$^3$ distribution;
 \item \dots
 \item $p=\prod_{k=0}^K p_k$, the exact target distribution, which is also the tilted$^K$ distribution.
 \end{itemize}
From a variational perspective, expressive approximating families for $g$, that is, beyond exponential families, could be used to improve the individual site approximations
\citep{tran2016variational,ranganath2016hierarchical}.
Instead of independent groups, tree structures could also be used~\citep{Opper+Winther:2005}. Even something as simple as mixing simulation draws from the tilted distribution could be a reasonable improvement on its approximation. One could then go further, for example at convergence computing simulations from some of the tilted distributions.

Message passing algorithms can be combined with other approaches to data partitioning.  In the present paper, we have focused on the construction of the approximate densities $g_k$ with the goal of simply multiplying them together to get the final approximation $g=p_0\prod_{k=1}^Kg_k$. However, one could instead think of the cavity distributions $g_{-k}$ at the final iteration as separate priors, and then follow the ideas of \citet{Wang+Dunson:2013}.

Another direction is to compare the global approximation with the tilted distribution, for example by computing a Kullback-Leibler divergence or looking at the distribution of importance weights.  Again, we can compute all the densities analytically, we have simulations from the tilted distributions, and we can trivially draw simulations from the global approximation, so all these considerations are possible.

%%% Local Variables:
%%% mode: latex
%%% TeX-master: "ep_statsci"
%%% End:

\section{Experiments}\label{experiments}

As discussed in Section~\ref{stuff}, the distributed EP framework can be applied to problems in various ways. In this section, we implement an algorithm using MCMC for tilted distribution inference, demonstrating in two hierarchical examples: a simulated logistic regression problem and a mixture model applied to astronomy data.
More details of the experiments can be found
in Appendix~\ref{appendix:implementational_details}.

The objective of these experiments is to demonstrate the EP framework as a convenient method for distributing inference carried out by general probabilistic programming tools. These experiments do not serve as a thorough examination of the principles of EP in itself or as an exhaustive comparison between competitive distributed inference algorithms.

% Simulated Example
\subsection{Simulated hierarchical logistic regression}\label{subsec:simulated_data_experiment}

We demonstrate the distributed EP algorithm with a simulated hierarchical logistic regression problem, a typical case in statistical analysis.
We inspect the behavior of the method when increasing the number of partitions, which is expected to speed up the inference but decrease the approximation accuracy.
We compare to consensus Monte Carlo~\citep{Scott+others:2016}, an alternative distributed sampling method.
Non-sampling based methods are not used as a comparison. In particular, related but non-distributable conventional optimization based variational inference is not analyzed here.
We consider full non-distributed MCMC approximation as a reference method.
The aim of the experiment is to show that the method is applicable and that it can outperform consensus Monte Carlo~\citep{Scott+others:2016}.

In the context of distributed computing, the constructed problem is small with 64 groups and 1280 observations in total. When comparing to the non-distributed inference, the gains in the computational efficiency should be greater with bigger problems.
The time complexity of the distributed algorithm is mostly determined by the MCMC sampling in the local sites.
Thus the expected complexity simplifies to $\operatorname{O}(h(n/K, d_\phi))$, where $n$ is the total number of observations, $K$ is the number of sites, $d_\phi$ is the dimensionality of the shared parameters, and $h(n/K, d_\phi)$ indicates the complexity of sampling the local model with $n/K$ observations and dimensionality $d_\phi$.
More detailed analysis of the computational complexity is presented in Appendix~\ref{appendix_cost}.
In general, memory efficiency and limitations must also be taken into consideration, as the dataset might not fit in the memory in the first place and then it would need to be partitioned.

The problem has not been chosen here with the expectation that it would be particularly easy to approximate with the method. On the contrary, it can be seen from the results that unlike in the non-hierarchical logistic regression, where EP is known to perform well, the hierarchical problem hard as EP tends to underestimate the variance when there are many sites and strong posterior dependencies~\citep{Cunningham+others:2011,cseke2013approximate}.

The model we shall fit is
\begin{align*}
    \left. y_{ij} \kern 1pt \middle| \kern 1pt x_{ij}, \beta_j \right.
        &\sim \operatorname{Bernoulli}\Bigl(\operatorname{logit}^{-1}\bigl(f_{ij}\bigr)\Bigr),
\intertext{where}
    f_{ij} &= \beta_j^\mathrm{T} x_{ij} \\
    \beta_{jd} &\sim \mbox{N}\bigl(\mu_d,\sigma_d^2\bigr),\\
    \mu_d &\sim \mbox{N}\bigl(0,\tau_\mu^2\bigr),\\
    \sigma_d &\sim \mbox{log-N}\bigl(0,\tau_\sigma^2 \bigr),
\end{align*}
for all dimensions $d = 0,1,\dots,D$, groups $j = 1,2,\dots,J$, and observations $i = 1,2,\dots,n_j$. The observed data have $D$ features. The first coefficient $\beta_0$ corresponds to the intercept; correspondingly, the first element in the data vector $x_{i,j}$ is a column of 1's. The shared parameters inferred with EP are $\phi = (\mu, \, \log\sigma)$.  Figure~\ref{fig:experiment_model} shows the structure of the model.

\begin{figure}
\centering
%~ \documentclass[border=10pt]{standalone}
%~ 
%~ % ========== Settings (start) ==================================================
%~ \usepackage[T1]{fontenc}
%~ \usepackage[utf8]{inputenc}
%~ \usepackage{lmodern}
%~ % \usepackage{mathtools}
%~ \usepackage{isomath}
%~ \newcommand*{\+}[1]{\ensuremath{\vectorsym{#1}}}
%~ % ========== Settings (end) ====================================================
%~ 
%~ \usepackage{tikz}
%~ \usetikzlibrary{bayesnet}
%~ 
%~ \begin{document}

\begin{tikzpicture}

  % Define nodes
  \node[obs]                               (x) {$x_{ij}$};
  \node[latent, right=of x]                (f) {$f_{ij}$};
  \node[obs, right=of f]                   (y) {$y_{ij}$};
  
  \node[latent, above=6mm of f] (b) {$\beta_j$};
  \node[latent, above=6mm of b, xshift=-10mm] (mb) {$\mu$};
  \node[latent, above=6mm of b, xshift= 10mm] (sb) {$\sigma$};
  
  \node[const, above=6mm of sb] (ssb0) {$\tau_{\sigma}$};
  \node[const, above=6mm of mb] (smb0) {$\tau_{\mu}$};
    
  % Connect the nodes
  \edge {x,b} {f};
  \edge[overlay] {f} {y};
  \edge {mb,sb} {b};
  \edge {ssb0} {sb};
  \edge {smb0} {mb};

  % Plates
  \plate {xfy} {(x)(f)(y)} {$i=1,2,\dots,N_j$} ;
  \plate {groups} {(xfy.south west)(xfy.south east)(b)} {$j=1,2,\dots,J$} ;

\end{tikzpicture}

%~ \end{document}
\caption{
A graphical model representation of the experimented hierarchical logistic regression problem. Indexing $j=1,2,\dots,J$ corresponds to hierarchical groups and $i=1,2,\dots,n_j$ corresponds to observations in group $j$. Gray nodes represent observed variables and white nodes represent unobserved latent variables. Variables without circles denote fixed priors.}\label{fig:experiment_model}
\end{figure}

The simulated problem is constructed with a $D=16$ dimensional explanatory variable resulting in a total of $d_\phi=2(D+1)=34$ shared parameters. The number of hierarchical groups is $J=64$ and the number of data points per group is $n_j=20$ for all $j=1,\ldots,J$, resulting in a total of $N=1280$ data points.
The correlated explanatory variable is sampled from a normal distribution $\mbox{N}\bigl(\mu_{x_j},\Sigma_{x_j}\bigr)$, where $\mu_{x_j}$ and $\Sigma_{x_j}$ are regulated so that the latent probability $\operatorname{logit}^{-1}\bigl(\beta_j^\mathrm{T} x_{ij}\bigr)$ gets values near 0 and 1 with low but not too low frequency.
This ensures that the problem is neither too easy nor too hard.
We present the details of the regularization
in Appendix~\ref{appendix:ex1_details}.

\begin{figure}
\centering
\input{figures/fig_ex1_timex.pgf}
\caption{
MSE of the mean and approximate KL divergence from the target distribution to the resulting posterior approximation as a function of the elapsed sampling time.
%Time spend in non-sampling parts of the code is negligible.
Three methods are compared: full MCMC, distributed EP, and distributed consensus MC.
For EP (solid lines) and consensus Monte Carlo (dotted lines), line colors indicate the number of partitions $K$.
The $y$-axis is in the logarithmic scale.
Unsurprisingly, the final accuracy declines as the number of partitions increases.
In all partitionings, EP outperforms consensus MC, and with small $K$, it reaches comparable accuracy to the full MCMC approximation. The sampling time comparison is tentative, as the EP implementation could be further optimized by reusing sampling parameters in consecutive iterations. In addition to the time efficiency, the reduced memory usage per distributed unit, gained by increasing the number of partitions $K$, would also be a concern for large problems.}\label{fig:experiment_mse}
\end{figure}

\begin{figure}[tb]
\centering
\input{figures/fig_ex1_kx.pgf}
\caption{
The resulting MSE and KL divergence with distributed EP and consensus Monte Carlo as a function of the number of partitions. EP reached better results in all cases. The $y$-axis is in the base-10 logarithmic scale and the $x$-axis is in the base-2 logarithmic scale.}\label{fig:experiment_ex1_ks}
\end{figure}

\begin{figure}[tb]
\centering
\input{figures/fig_ex1_comp.pgf}
\caption{
Pointwise comparison of the posterior mean and standard deviation of the target and the final distributed EP approximation when the groups are distributed into $K=2$ (top row) and $K=64$ (bottom row) sites. Each dot corresponds to one of the 34 shared parameters. The red diagonal line shows the points of equivalence. It can be seen that in this experiment, while accurately finding the mean, EP systematically underestimates the variance.}\label{fig:experiment_compare}
\end{figure}

Following the hierarchical EP algorithm description in Section~\ref{hier}, we run experiments partitioning the data into $K = 2,4,8,16,32,64$ sites, using uniform distributions as initial site approximations.
Distributing the problem further into $K>64$ sites, and ultimately to $K = \sum_{j=1}^J n_j$ sites corresponding to the conventional fully factored EP, would require that the local parameters be included in the global EP approximation, thus loosing the advantage of the hierarchical setting. This drastic increase in the shared parameter space would often make the approach inapplicable and thus we omit these experiments here.

Our implementation uses Python for the message passing framework and the Stan probabilistic modeling language~\citep{Stan:2017} for MCMC sampling from the tilted distribution.
The tilted distribution moments are estimated in natural form with~\eqref{eq:n_sample_prec_estim_Q} and~\eqref{eq:n_sample_prec_estim_r} from the obtained sample.
Each parallel MCMC run has 8 chains of length 200, in which the first halves of the chains are discarded as warmup.
In our implementation, the warmup period, during which sampling parameters are learned, is performed in every iteration of EP. It would be possible, however, to adopt the state of the sampler from previous iteration to speed up the process.
As discussed before in Section~\ref{sec_damping}, we apply gradually decreasing damping factor $\delta$.
In our experiment, the following setup produced good results; in the first iteration, $\delta = 0.5$ and it decays exponentially towards $\min(1/K, 0.2)$ while reaching 90 \% decay at iteration $K$.

We compare the results from the distributed EP approximations to a distributed consensus Monte Carlo approximation~\citep{Scott+others:2016} and undistributed full MCMC approximation with varying sample size. In the consensus method, the data is split analogously to $K = 2,4,8,16,32,64$ partitions and the prior is respectively fractioned to $p(\theta)^{1/K}$ in each separate inference.
All of the obtained results are compared to a target full MCMC approximation with 8 chains of length $10\,000$, in which the first halves of the chains are discarded as warmup.
The code for the experiments is available at \url{https://github.com/gelman/ep-stan}.

If we were to use a simple scheme of data splitting and separate inferences (without using the cavity distribution as an effective prior distribution at each step), the computation would be problematic:  with only 20 data points per group, each of the local posterior distributions would be wide, as sketched in Figure~\ref{sketch}.  The message passing framework, in which at each step the cavity distribution is used as a prior, keeps computations more stable and focused.

Figure~\ref{fig:experiment_mse} illustrates the progression of the experiment for each run.
In this experiment, EP reached better accuracy than consensus MC with all $K$.
As shown
in Appendix~\ref{appendix:ex1_extra_results},
the difference in the accuracy between EP and consensus MC becomes bigger, if the parameter correlations in the KL divergence measure are ignored.
With small $K$, EP was able to reach accuracy comparable to the full sampling.
Figure~\ref{fig:experiment_ex1_ks} compares the final obtained approximation accuracy between EP and consensus method with varying $K$.
In both of these methods, the final approximation quality is better with fewer sites but more sites provide opportunities for faster convergence and reduced memory usage per unit in parallelized setting.
Figure~\ref{fig:experiment_compare} shows a comparison between posterior mean and standard deviation between the distributed EP approximation and the target approximation for the shared parameters in the extreme cases $K=2$ and $K=64$.
Points closer to the red diagonal line imply a better EP approximation. It can be seen that the case $K=2$ results in an overall better approximation.

As discussed before in the start of this section, it can be seen from Figure~\ref{fig:experiment_compare} that EP tends to underestimate the variance with more sites. This underestimation is a known feature in EP when there are many sites and strong posterior dependencies~\citep{Cunningham+others:2011, cseke2013approximate}. However, unlike with the consensus MC method, the mean is well approximated with distributed EP even with a high number of partitions, as can be seen from Figure~\ref{fig:experiment_mse}.

Although not the focus of this experiment, in Figure~\ref{fig:experiment_mse}, we assess the time efficiency of the method by inspecting the performance indicator as a function of the time spent in the sampling parts of the code. By this, we can compare the methods in an even manner by neglecting the implementation-specific factor. Each of the methods uses the same Stan implementation for the sampling.
In our experiments, the time spent in other parts of the code is minuscule compared to the sampling time; even in the most extreme case of $K=64$, the time spent in non-sampling parts of the code was only 0.2\% of the total time spent.
However, as various aspects of the problem affect the computational efficiency, our general time comparison is tentative. For example, it should be possible to improve the sampling time in EP by adopting the sampling parameters from previous EP iterations. %We did not implement this because Stan does not support this at the moment.
We further discuss the computational complexity of the method in Appendix~\ref{appendix_cost}.

%Astronomy Example
\subsection{Hierarchical mixture model applied to an astronomy problem}\label{subsec:astronomy_data_experiment}

We next apply the distributed EP algorithm to a problem in astronomy, where the goal is to model the nonlinear relationship between diffuse galactic far ultraviolet radiation (FUV) and 100-$\mu$m infrared emission (i100) in various sectors of the observable universe. The data were collected from the Galaxy Evolution Explorer telescope. An approximate linear relationship has been found between FUV and i100 below i100 values of 8 MJy sr$^{-1}$~\citep{Hamden+others:2013}.  Here we attempt to model the nonlinear relationship across the entire span of i100 values, allowing the curves to vary spatially.
\citet{Sahai:2018phd} discusses the experiment in more detail and also presents some additional simulated experiments using the same method.

\begin{figure}[htbp]
\centering
\input{figures/fig_astro_data_arxiv.tex}
\caption{
Scatterplots of far ultraviolet radiation (FUV) versus infrared radiation (i100) in various regions of the universe. Data are shown for regions of longitude $12\degree, 23\degree, 92\degree$, and $337\degree$, and are presented with axes on the original scale (first column) and on the log scale (second column).}\label{fig:astro_data}
\end{figure}

Figure~\ref{fig:astro_data} shows scatterplots of FUV versus i100 in different longitudinal regions (each of width 1 degree) of the observable universe. The bifurcation in the scatterplots for i100 values greater than 8 MJy sr$^{-1}$ suggests a nonlinear mixture model is necessary to capture the relationship between the two variables. At the same time, a flexible parametric model is desired to handle the various mixture shapes, while maintaining interpretability in the parameters.

\begin{figure}[tb]
\centering
   \includegraphics{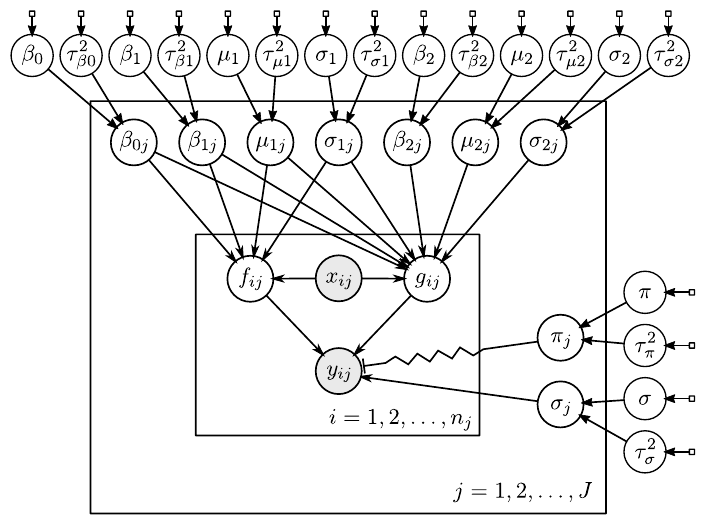}
\caption{
Graphical representation of the astronomy model. Circles represent random variables and boxes represent fixed parameters. Grayed circles are observed. The zig-zag line indicates that $\pi_j$ functions as a selector between $f_{ij}$ and $g_{ij}$. The labels for the fixed prior parameters are omitted for clarity.
}\label{fig:astro_model}
\end{figure}

Letting $\sigma(\cdot) = \text{logit}^{-1}(\cdot)$ denote the inverse logistic function and letting
\begin{equation*}
a_j = \left( \beta_{0j}, \beta_{1j}, \mu_{1j}, \sigma_{1j}, \sigma^{-1}(\beta_{2j}), \mu_{2j}, \sigma_{2j}, \sigma^{-1}(\pi_j), \sigma_j \right)
\end{equation*}
denote the local parameters for each group $j$, we model the top part of the bifurcation (the first component of the mixture) as a generalized inverse logistic function,
$$ f(a_j, x_{ij}) = \beta_{0j} + \beta_{1j} \sigma \biggl( \frac{ \log x_{ij} - \mu_{1j} }{ \sigma_{1j} } \biggr), $$
while the second mixture component is modeled as the same inverse logistic function multiplied by an inverted Gaussian:
$$ g(a_j, x_{ij}) = \beta_{0j} + \beta_{1j} \sigma \biggl( \frac{ \log x_{ij} - \mu_{1j} }{ \sigma_{1j} } \biggr) \cdot
\biggl( 1 - \beta_{2j} \exp \biggl( -\frac{1}{2} \bigg( \frac{ \log x_{ij} - \mu_{2j} }{ \sigma_{2j} } \biggr)^2 \biggr) \biggr). $$
As such, the ultraviolet radiation ($y_{ij}$) is modeled as a function of infrared radiation ($x_{ij}$) through the following mixture model:
\begin{align*}
    \log y_{ij} &= \pi_j \cdot f(a_j, x_{ij}) + (1 - \pi_j) \cdot g(a_j, x_{ij}) + \sigma_j \epsilon_{ij},\\
    \epsilon_{ij} &\sim N(0,1),
\intertext{where $\beta_{2j} \in [0,1]$, $\pi_j \in [0,1]$, and the local parameters are modeled hierarchically with the following shared centers and scales: }
    \beta_{0j} &\sim \mbox{N}\bigl(\beta_0,\tau_{\beta 0}^2),\\
    \beta_{1j} &\sim \mbox{N}\bigl(\beta_1,\tau_{\beta 1}^2),\\
    \mu_{1j} &\sim \mbox{log-N}\bigl(\log \mu_1,\tau_{\mu 1}^2),\\
    \sigma_{1j} &\sim \mbox{log-N}\bigl(\log \sigma_1,\tau_{\sigma 1}^2\bigr),\\
    \sigma^{-1}(\beta_{2j}) &\sim \mbox{N}\bigl(\sigma^{-1}(\beta_2),\tau_{\beta 2}^2 \bigr),\\
    \mu_{2j} &\sim \mbox{log-N}\bigl(\log \mu_2,\tau_{\mu 2}^2 \bigr),\\
    \sigma_{2j} &\sim \mbox{log-N}\bigl(\log \sigma_2,\tau_{\sigma 2}^2\bigr),\\
    \sigma^{-1}(\pi_j) &\sim \mbox{N}\bigl(\sigma^{-1}(\pi),\tau_\pi^2 \bigr),\\
    \sigma_j &\sim \mbox{log-N}\bigl(\sigma,\tau_\sigma^2 \bigr)
\end{align*}
for all groups $j = 1,2,\dots,J$, and observation $i = 1,2,\dots,n_j$. The model is illustrated graphically in Figure~\ref{fig:astro_model}.

Hence the problem has $9 \cdot 2 = 18$ shared parameters of interest. The number of local parameters depends on how finely we split the data in the observable universe. Our study in particular is constructed with $J = 360$ hierarchical groups (one for each longitudinal degree of width one degree), resulting in a total of $9  J = 3\,240$ local parameters. We also sample the number of observations per group as $n_j = 2\,000$ for all $j = 1,\ldots,J$, resulting in a total of $N=720\,000$ observations.

When dividing the longitudal degrees into distinct hierarchical groups, the relative angular distance between groups is ignored; nearby groups are considered equally dependent as far away ones.
This is often an issue with divide-and-conquer algorithms when the data have spatial or temporal structure.
Increasing the number of partitions ignores more information but also increases computational efficiency.
In addition, one must pay attention to local coherence in the groupings.
We find that applying this model for the problem is reasonable, and it also serves as an example for hierarchical nonlinear regressions more generally.
\citet{Sahai:2018phd} discusses the matter in more detail.

Our implementation uses R for the message passing framework and the Stan probabilistic modeling language~\citep{Stan:2017} for MCMC sampling from the tilted distribution. We fit the mixture model with various EP settings, partitioning the data into $K = 5, 10, 30$ sites and using uniform distributions as the initial site approximations. For the tilted distribution inference, the natural parameters are estimated using~\eqref{eq:n_sample_prec_estim_Q} and~\eqref{eq:n_sample_prec_estim_r}. Each parallel MCMC run has 4 chains with 1000 iterations each, of which half are discarded as warmup. We use a constant damping factor of $\delta=0.1$ in order to get coherent convergence results amongst different partitions. We compare the results from the distributed EP approximations to an MCMC approximation for the full model using Stan. The full approximation uses 4 chains with 1000 iterations each, of which half are discarded as warmup.

\begin{figure}[tbp]
\centering
   \includegraphics[width=0.82\textwidth]{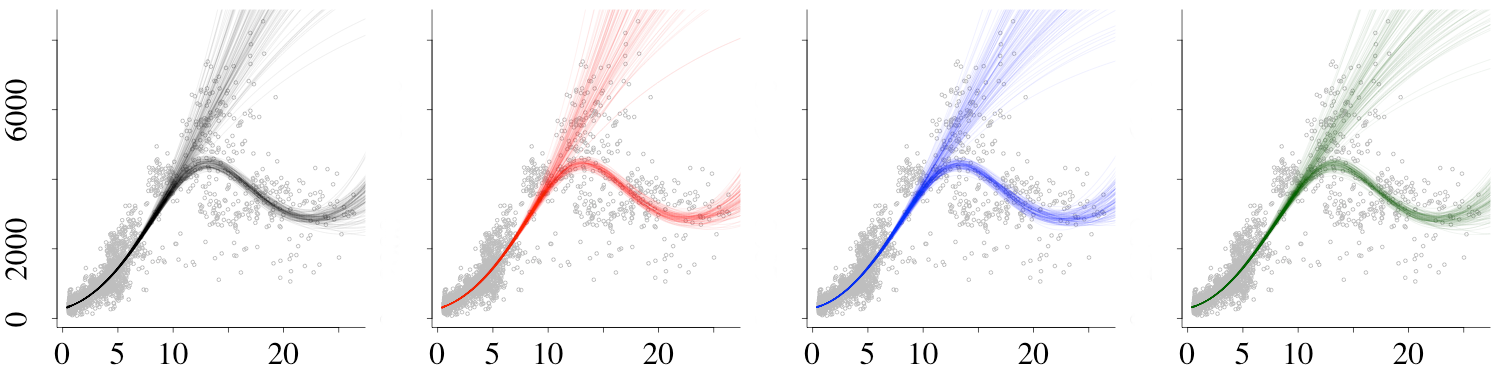}
   \includegraphics[width=0.82\textwidth]{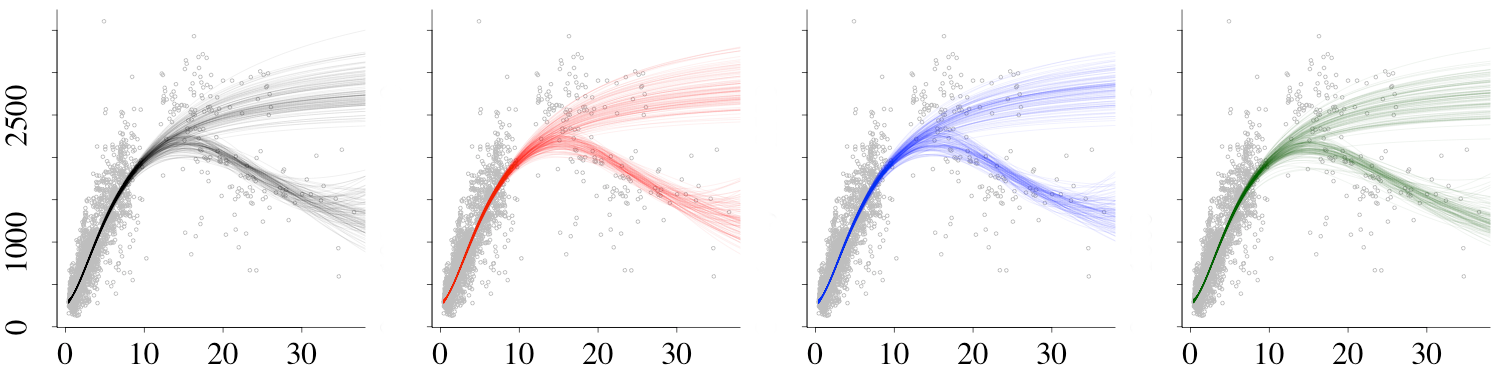}
   \includegraphics[width=0.82\textwidth]{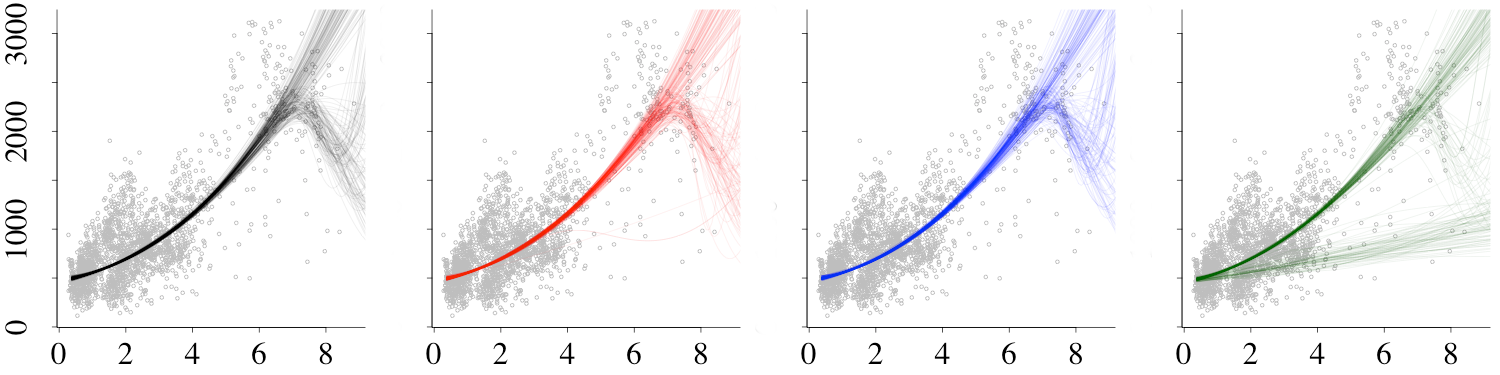}
   \includegraphics[width=0.82\textwidth]{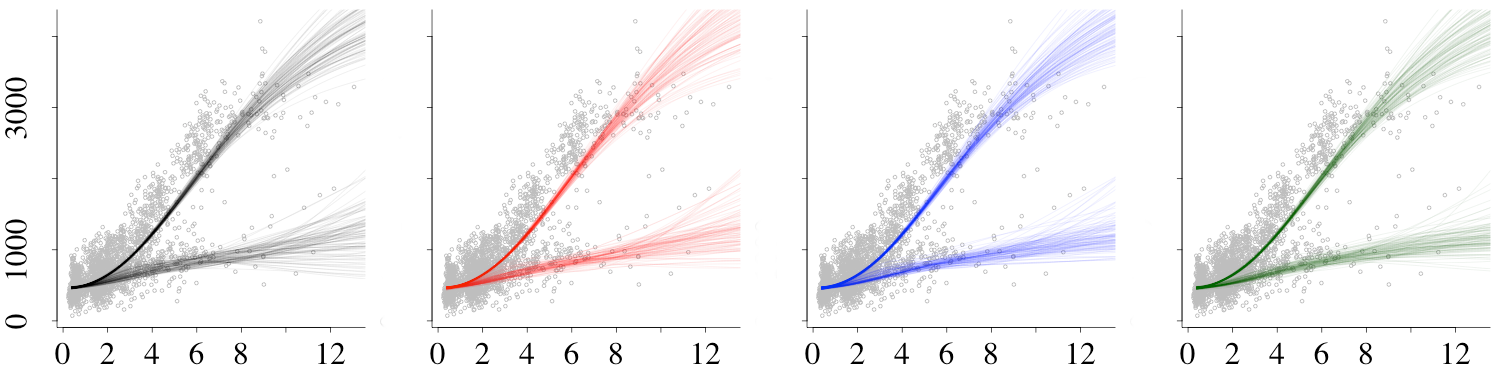}
   \includegraphics[width=0.82\textwidth]{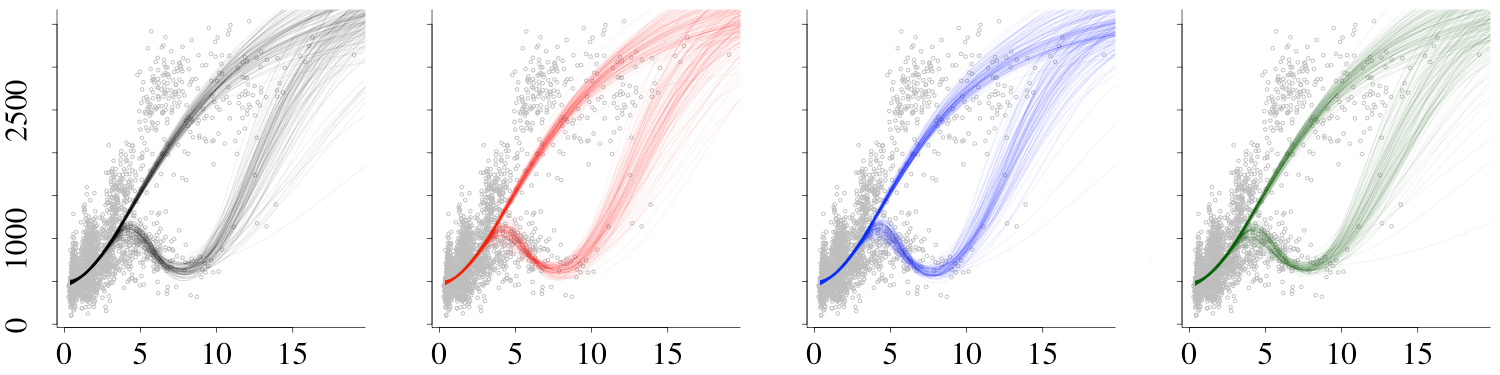}
   \includegraphics[width=0.82\textwidth]{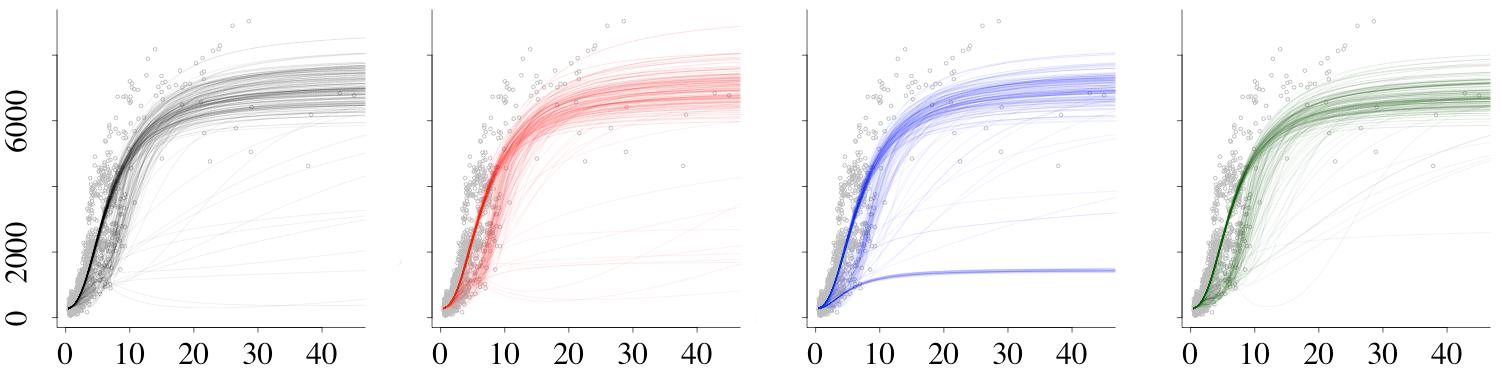}
\caption{
Comparison of the local fits of the full MCMC computation (black) for the astronomy example and the final distributed EP approximations when the groups are distributed into $K=5$ (red), $K=10$ (blue), and $K=30$ (green) sites. Posterior draws are shown for each of 6 groups (one group per row) with longitudes $12\degree, 32\degree, 82\degree, 92\degree, 93\degree,$ and $194\degree$.}\label{fig:astro_compare}
\end{figure}

\begin{figure}[tb]
\centering
   \includegraphics[width=0.6\textwidth]{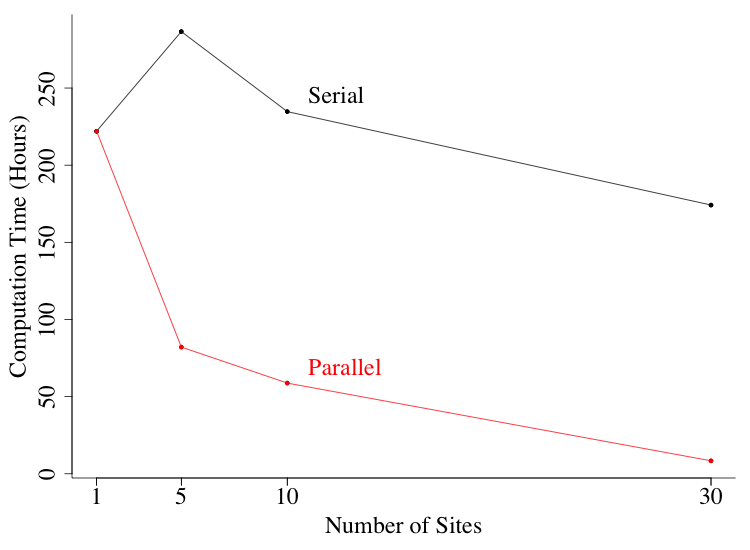}
\caption{
Computation times for the distributed EP algorithm applied to the astronomy data, as a function of the number of sites. The full MCMC computation time is equivalent to that of EP with $K=1$ site. The computational benefits of increasing the number of sites is clear when the updates are parallel.}\label{fig:astro_times}
\end{figure}

Figure~\ref{fig:astro_compare} shows a comparison of the local scatterplot fits for each EP setting on various hierarchical groups, each representing a one-degree longitudinal slice of the observable universe. While all of the runs show similar results for most groups, there are some cases where increasing the number of sites results in poorer performance. In particular, EP with 30 sites converges to a different mixture for $82\degree$, while EP with 10 sites converges to a different mixture for $194\degree$.

Figure~\ref{fig:astro_times} illustrates the computation times for the EP runs with serial and parallel updates. The advantages of distributed EP are most clear when comparing K = 1 site to K = 30 sites,  which results in a 96\% decrease in computation time. This advantage in computation time, however, depends on the implementation of the parallelization. By using the time spent on the sampling of the tilted distribution as our benchmarking criterion, we can focus on the crucial part of the algorithm and neglect the implementation-specific factor.

%%% Local Variables:
%%% mode: latex
%%% TeX-master: "ep_statsci"
%%% End:

\section{Discussion}\label{discussion}

Using the principle of message passing with cavity and tilted distributions, we have presented a framework for Bayesian inference on partitioned data sets.
Similar to more conventional divide-and-conquer algorithms, EP can be used to divide the computation into manageable sizes without scattering the problem into too small pieces.
Furthermore, EP comes with the additional advantage of naturally sharing information
between distributed parts, focusing the computation into important areas of the parameter space. In our experiment, the method outperforms comparable consensus MC algorithm~\citep{Scott+others:2016} both in time and approximation error.

Probabilistic programming languages such as Stan~\citep{Stan:2017} provide convenient generalizable tools for statistical data analysis. When dealing with problems where the data does not fit in the memory, the EP framework can be included in the process to distribute the computation without loosing the generalisability.
>From an alternative point of view, EP can also be used to pool information
across many sources of already partitioned data sets and models.
In the case of hierarchical models, EP enables efficient distributed computation for large models with big data sets, as well as meta-models fit into local models or local aggregated data.

The message passing framework presented in this paper includes
numerous design choices, and many methods can be subsumed under it.
In particular, the inference in the sites can be implemented in various ways.
This extensive configurability provides possibilities for improved efficiency but also makes it more complex to set up.
In this paper we have discussed two generalizable simulation based methods in particular, moment matching and SNEP~\citep{Hasenclever+others:2017}.
These methods perform better in different situations. It is also possible to use them in combination, for example  using moment matching in early iterations for a more stable and quicker start, and SNEP in later iterations for more stable and precise convergence.
If convergence problems are encountered, while different choices in the method may also be helpful, it could be useful to first consider alternative approximating families instead.
Further research is required in order to learn the effect of different configurations and the optimal approaches to various problem settings.

Data partitioning is an extremely active research area with several
black box algorithms being proposed by various research groups~\citep[e.g.][]{Kucukelbir+others:2016,Hasenclever+others:2017,bardenet2015on}.
We are sure that different methods will be more effective in different problems.
The present paper has two roles: we review the steps that are needed to keep EP algorithms numerically stable, and we are suggesting a general approach, inspired by EP, for approaching data partitioning problems in a way that achieves the computational benefits of parallelism while allowing each local update to make use of relevant information from the other sites.

While EP may not yet be a ``way of life,`` we argue that the increasing popularity of divide-and-conquer algorithms in big data environments is moving us in this direction.
Stepping back from particular choices in implementation, the idea of the cavity and tilted distributions seems to us to be crucial in understanding how inferences from separate pieces of information can be combined in a way that respects the model being fit. We anticipate that great progress could be made by using message passing to regularize existing algorithms.

\section*{Acknowledgements}
We thank David Blei, Ole Winther, Bob Carpenter, and anonymous reviewers for helpful comments for helpful comments, and the U.S. National Science Foundation, Institute for Education Sciences, Office of Naval Research, Moore and Sloan Foundations, and Academy of Finland (grant 298742, 313122, and the Finnish Centre of Excellence in Computational Inference Research COIN) for partial support of this research. We also acknowledge the computational resources provided by the Aalto Science-IT project.

\bibliography{ep}

\begin{thebibliography}{75}
\providecommand{\natexlab}[1]{#1}
\providecommand{\url}[1]{\texttt{#1}}
\expandafter\ifx\csname urlstyle\endcsname\relax
  \providecommand{\doi}[1]{doi: #1}\else
  \providecommand{\doi}{doi: \begingroup \urlstyle{rm}\Url}\fi

\bibitem[Ahn et~al.(2012)Ahn, Korattikara, and Welling]{Ahn+others:2012}
Sungjin Ahn, Anoop Korattikara, and Max Welling.
\newblock Bayesian posterior sampling via stochastic gradient {F}isher scoring.
\newblock In \emph{Proceedings of the 29th International Conference on Machine
  Learning}, 2012.

\bibitem[Balan et~al.(2014)Balan, Chen, and Welling]{Korattikara+others:2014}
Anoop~Korattikara Balan, Yutian Chen, and Max Welling.
\newblock Austerity in {MCMC} land: Cutting the {M}etropolis-{H}astings budget.
\newblock In \emph{Proceedings of the 31th International Conference on Machine
  Learning, {ICML} 2014, Beijing, China, 21-26 June 2014}, pages 181--189,
  2014.

\bibitem[Bardenet et~al.(2015)Bardenet, Doucet, and Holmes]{bardenet2015on}
R{\'e}mi Bardenet, Arnaud Doucet, and Chris Holmes.
\newblock {On Markov chain Monte Carlo methods for tall data}.
\newblock \emph{arXiv.org}, 2015.

\bibitem[Barthelm\'e and Chopin(2014)]{Barthelme+Chopin:2014}
Simon Barthelm\'e and Nicolas Chopin.
\newblock Expectation propagation for likelihood-free inference.
\newblock \emph{Journal of the American Statistical Association}, 109:\penalty0
  315--333, 2014.

\bibitem[Betancourt(2013)]{Betancourt:2013}
Michael Betancourt.
\newblock A general metric for {R}iemannian manifold {H}amiltonian {M}onte
  {C}arlo.
\newblock In Frank Nielsen and Fr{\'e}d{\'e}ric Barbaresco, editors,
  \emph{Geometric Science of Information - First International Conference},
  pages 327--334, Berlin, Heidelberg, 2013. Springer.

\bibitem[Betancourt(2014)]{Betancourt:2014}
Michael Betancourt.
\newblock Adiabatic {M}onte {C}arlo.
\newblock \emph{arXiv preprint arXiv:1405.3489}, 2014.

\bibitem[Bodnar and Gupta(2011)]{Bodnar+Gupta:2011}
Taras Bodnar and Arjun~K. Gupta.
\newblock Estimation of the precision matrix of a multivariate elliptically
  contoured stable distribution.
\newblock \emph{Statistics}, 45\penalty0 (2):\penalty0 131--142, 2011.

\bibitem[Bodnar et~al.(2014)Bodnar, Gupta, and Parolya]{Bodnar+others:2014}
Taras Bodnar, Arjun~K. Gupta, and Nestor Parolya.
\newblock Optimal linear shrinkage estimator for large dimensional precision
  matrix.
\newblock \emph{arXiv preprint arXiv:1308.0931}, 2014.

\bibitem[Chen and Wand(2018)]{chen_wand2018factor_graph_fragments}
Wilson~Y. Chen and Matt~P. Wand.
\newblock Factor graph fragmentization of expectation propagation.
\newblock \emph{arXiv:1801.05108}, 2018.

\bibitem[Chib(1995)]{Chib:1995}
Siddhartha Chib.
\newblock Marginal likelihood from the {G}ibbs output.
\newblock \emph{Journal of the American Statistical Association}, 90\penalty0
  (432):\penalty0 1313--1321, 1995.

\bibitem[Cornuet et~al.(2012)Cornuet, Marin, Mira, and
  Robert]{Cornuet+others:2012}
Jean-Marie Cornuet, Jean-Michel Marin, Antonietta Mira, and Christian~P.
  Robert.
\newblock Adaptive multiple importance sampling.
\newblock \emph{Scandinavian Journal of Statistics}, 39\penalty0 (4):\penalty0
  798--812, 2012.

\bibitem[Cseke and Heskes(2011)]{Cseke+Heskes:2011}
Botond Cseke and Tom Heskes.
\newblock Approximate marginals in latent {Gaussian} models.
\newblock \emph{Journal of Machine Learning Research}, 12:\penalty0 417--454,
  2011.

\bibitem[Cseke et~al.(2013)Cseke, Opper, and Sanguinetti]{cseke2013approximate}
Botond Cseke, Manfred Opper, and Guido Sanguinetti.
\newblock Approximate inference in latent {Gaussian-Markov} models from
  continuous time observations.
\newblock In \emph{Advances in Neural Information Processing Systems}, pages
  971--979, 2013.

\bibitem[Cunningham et~al.(2011)Cunningham, Hennig, and
  Lacoste-Julien]{Cunningham+others:2011}
John~P. Cunningham, Philipp Hennig, and Simon Lacoste-Julien.
\newblock Gaussian probabilities and expectation propagation.
\newblock \emph{arXiv preprint arXiv:1111.6832}, 2011.

\bibitem[Dean et~al.(2012)Dean, Corrado, Monga, Chen, Devin, Mao, Ranzato,
  Senior, Tucker, Yang, Le, and Ng]{dean2012large}
Jeffrey Dean, Greg Corrado, Rajat Monga, Kai Chen, Matthieu Devin, Mark Mao,
  Marc'aurelio Ranzato, Andrew Senior, Paul Tucker, Ke~Yang, Quoc~V. Le, and
  Andrew~Y. Ng.
\newblock {Large Scale Distributed Deep Networks}.
\newblock In \emph{Neural Information Processing Systems}, pages 1223--1231,
  2012.

\bibitem[Dehaene(2016)]{dehaene2016expectation}
Guillaume Dehaene.
\newblock Expectation propagation performs a smoothed gradient descent.
\newblock \emph{arXiv preprint arXiv:1612.05053}, 2016.

\bibitem[Dehaene and Barthelm{\'e}(2018)]{dehaene2018expectation}
Guillaume Dehaene and Simon Barthelm{\'e}.
\newblock Expectation propagation in the large data limit.
\newblock \emph{Journal of the Royal Statistical Society: Series B (Statistical
  Methodology)}, 80\penalty0 (1):\penalty0 199--217, 2018.

\bibitem[Dehaene and Barthelm\'{e}(2015)]{dehaene2015bounding}
Guillaume~P Dehaene and Simon Barthelm\'{e}.
\newblock Bounding errors of expectation-propagation.
\newblock In \emph{Advances in Neural Information Processing Systems 28}, pages
  244--252. Curran Associates, Inc., 2015.

\bibitem[Del~Moral et~al.(2006)Del~Moral, Doucet, and
  Jasra]{DelMoral+others:2006}
Pierre Del~Moral, Arnaud Doucet, and Ajay Jasra.
\newblock {Sequential Monte Carlo samplers}.
\newblock \emph{Journal of the Royal Statistical Society B}, 68\penalty0
  (3):\penalty0 411--436, 2006.

\bibitem[Dominici et~al.(1999)Dominici, Parmigiani, Wolpert, and
  Hasselblad]{Dominici+others:1999}
Francesca Dominici, Giovanni Parmigiani, Robert~L. Wolpert, and Vic Hasselblad.
\newblock Meta-analysis of migraine headache treatments: Combining information
  from heterogeneous designs.
\newblock \emph{Journal of the American Statistical Association}, 94\penalty0
  (445):\penalty0 16--28, 1999.

\bibitem[Dwork and Roth(2014)]{Dwork+Roth:2014}
Cynthia Dwork and Aaron Roth.
\newblock The algorithmic foundations of differential privacy.
\newblock \emph{Found. Trends Theor. Comput. Sci.}, 9\penalty0 (3--4):\penalty0
  211--407, 2014.

\bibitem[Friedman et~al.(2008)Friedman, Hastie, and
  Tibshirani]{Friedman+others:2008}
Jerome Friedman, Trevor Hastie, and Robert Tibshirani.
\newblock Sparse inverse covariance estimation with the graphical lasso.
\newblock \emph{Biostatistics}, 9\penalty0 (3):\penalty0 432--441, 2008.

\bibitem[Gelman et~al.(1996)Gelman, Bois, and Jiang]{Gelman+others:1996}
Andrew Gelman, Frederic Bois, and Jiming Jiang.
\newblock Physiological pharmacokinetic analysis using population modeling and
  informative prior distributions.
\newblock \emph{Journal of the American Statistical Association}, 91:\penalty0
  1400--1412, 1996.

\bibitem[Gelman et~al.(2008)Gelman, Jakulin, Pittau, and
  Su]{Gelman+others:2008}
Andrew Gelman, Aleks Jakulin, Maria~Grazia Pittau, and Yu-Sung Su.
\newblock A weakly informative default prior distribution for logistic and
  other regression models.
\newblock \emph{Annals of Applied Statistics}, 2:\penalty0 1360--1383, 2008.

\bibitem[Gelman et~al.(2014{\natexlab{a}})Gelman, Carpenter, Betancourt,
  Brubaker, and Vehtari]{Gelman+others:2014}
Andrew Gelman, Bob Carpenter, Michael Betancourt, Marcus Brubaker, and Aki
  Vehtari.
\newblock Computationally efficient maximum likelihood, penalized maximum
  likelihood, and hierarchical modeling using {S}tan.
\newblock Technical report, Department of Statistics, Columbia University,
  2014{\natexlab{a}}.

\bibitem[Gelman et~al.(2014{\natexlab{b}})Gelman, Vehtari, Jyl{\"a}nki, Robert,
  Chopin, and Cunningham]{Gelman+others:2014b}
Andrew Gelman, Aki Vehtari, Pasi Jyl{\"a}nki, Christian Robert, Nicolas Chopin,
  and John~P. Cunningham.
\newblock Expectation propagation as a way of life.
\newblock \emph{arXiv preprint arXiv:1412.4869v1}, 2014{\natexlab{b}}.

\bibitem[Geweke(1989)]{Geweke:1989}
John Geweke.
\newblock {Bayesian inference in econometric models using Monte Carlo
  integration}.
\newblock \emph{Econometrica}, 57\penalty0 (6):\penalty0 1317--1339, 1989.

\bibitem[Gupta et~al.(2013)Gupta, Varga, and Bodnar]{Gupta+others:2013}
Arjun~K. Gupta, Tamas Varga, and Taras Bodnar.
\newblock \emph{Elliptically Contoured Models in Statistics and Portfolio
  Theory}.
\newblock Springer-Verlag, New York, 2 edition, 2013.

\bibitem[Hamden et~al.(2013)Hamden, Schiminovich, and
  Seibert]{Hamden+others:2013}
Erika~T. Hamden, David Schiminovich, and Mark Seibert.
\newblock The diffuse galactic far-ultraviolet sky.
\newblock \emph{The Astrophysical Journal}, 779\penalty0 (180):\penalty0 15,
  December 2013.

\bibitem[Hasenclever et~al.(2017)Hasenclever, Webb, Lienart, Vollmer,
  Lakshminarayanan, Blundell, and Teh]{Hasenclever+others:2017}
Leonard Hasenclever, Stefan Webb, Thibaut Lienart, Sebastian Vollmer, Balaji
  Lakshminarayanan, Charles Blundell, and Yee~Whye Teh.
\newblock Distributed {B}ayesian learning with stochastic natural gradient
  expectation propagation and the posterior server.
\newblock \emph{Journal of Machine Learning Research}, 18\penalty0
  (106):\penalty0 1--37, 2017.

\bibitem[Hernandez-Lobato and Hernandez-Lobato(2016)]{hernandez_hernandez_2016}
Daniel Hernandez-Lobato and Jose~Miguel Hernandez-Lobato.
\newblock Scalable {Gaussian} process classification via expectation
  propagation.
\newblock In Arthur Gretton and Christian~C. Robert, editors, \emph{Proceedings
  of the 19th International Conference on Artificial Intelligence and
  Statistics}, volume~51 of \emph{Proceedings of Machine Learning Research},
  pages 168--176, Cadiz, Spain, 2016.

\bibitem[Hern{\'a}ndez-Lobato et~al.(2016)Hern{\'a}ndez-Lobato, Li, Rowland,
  Bui, Hern{\'a}ndez-Lobato, and Turner]{Hernandez:2016}
Jos{\'e}~Miguel Hern{\'a}ndez-Lobato, Yingzhen Li, Mark Rowland, Thang Bui,
  Daniel Hern{\'a}ndez-Lobato, and Richard~E. Turner.
\newblock Black-box $\alpha$-divergence minimization.
\newblock In \emph{Proceedings of the 33rd International Conference on Machine
  Learning, {ICML} 2016}, 2016.

\bibitem[Heskes and Zoeter(2002)]{Heskes+Zoeter:2002}
Tom Heskes and Onno Zoeter.
\newblock Expectation propagation for approximate inference in dynamic
  {Bayesian} networks.
\newblock In \emph{Proceedings of the Eighteenth Conference on Uncertainty in
  Artificial Intelligence}, UAI'02, pages 216--223, San Francisco, CA, USA,
  2002. Morgan Kaufmann Publishers Inc.

\bibitem[Heskes et~al.(2005)Heskes, Opper, Wiegerinck, Winther, and
  Zoeter]{heskes2005}
Tom Heskes, Manfred Opper, Wim Wiegerinck, Ole Winther, and Onno Zoeter.
\newblock Approximate inference techniques with expectation constraints.
\newblock \emph{Journal of Statistical Mechanics: Theory and Experiment},
  2005\penalty0 (11):\penalty0 P11015, 2005.

\bibitem[Higgins and Whitehead(1996)]{Higgins+Whitehead:1996}
Julian P.~T. Higgins and Anne Whitehead.
\newblock Borrowing strength from external trials in a meta-analysis.
\newblock \emph{Statistics in Medicine}, 15\penalty0 (24):\penalty0 2733--2749,
  1996.

\bibitem[Hoffman et~al.(2013)Hoffman, Blei, Wang, and
  Paisley]{Hoffman+others:2013}
Matthew~D. Hoffman, David~M. Blei, Chong Wang, and John~William Paisley.
\newblock Stochastic variational inference.
\newblock \emph{Journal of Machine Learning Research}, 14\penalty0
  (1):\penalty0 1303--1347, 2013.

\bibitem[Jyl{\"a}nki et~al.(2011)Jyl{\"a}nki, Vanhatalo, and
  Vehtari]{Jylanki+others:2011}
Pasi Jyl{\"a}nki, Jarno Vanhatalo, and Aki Vehtari.
\newblock Robust {Gaussian} process regression with a {Student-$t$} likelihood.
\newblock \emph{Journal of Machine Learning Research}, 12:\penalty0 3227--3257,
  2011.

\bibitem[Kucukelbir et~al.(2016)Kucukelbir, Tran, Ranganath, Gelman, and
  Blei]{Kucukelbir+others:2016}
Alp Kucukelbir, Dustin Tran, Rajesh Ranganath, Andrew Gelman, and David~M.
  Blei.
\newblock Automatic differentiation variational inference.
\newblock \emph{arXiv preprint arXiv:1603.00788}, 2016.

\bibitem[Lewandowski et~al.(2009)Lewandowski, Kurowicka, and
  Joe]{lewandowski2009generating}
Daniel Lewandowski, Dorota Kurowicka, and Harry Joe.
\newblock Generating random correlation matrices based on vines and extended
  onion method.
\newblock \emph{Journal of multivariate analysis}, 100\penalty0 (9):\penalty0
  1989--2001, 2009.

\bibitem[Li et~al.(2015)Li, Hern{\'a}ndez-Lobato, and Turner]{li:2015}
Yingzhen Li, Jos{\'e}~Miguel Hern{\'a}ndez-Lobato, and Richard~E. Turner.
\newblock Stochastic expectation propagation.
\newblock In \emph{Advances in Neural Information Processing Systems}, pages
  2323--2331, 2015.

\bibitem[Minka(2001{\natexlab{a}})]{Minka:2001a}
Thomas~P. Minka.
\newblock \emph{A family of algorithms for approximate Bayesian inference.}
\newblock PhD thesis, Massachusetts Institute of Technology, Cambridge, MA,
  USA, 2001{\natexlab{a}}.

\bibitem[Minka(2001{\natexlab{b}})]{Minka:2001b}
Thomas~P. Minka.
\newblock Expectation propagation for approximate {B}ayesian inference.
\newblock In J.~Breese and D.~Koller, editors, \emph{Proceedings of the 17th
  Conference in Uncertainty in Artificial Intelligence (UAI-2001)}, pages
  362--369. Morgan Kaufmann, San Francisco, Clif., 2001{\natexlab{b}}.

\bibitem[Minka(2004)]{Minka:2004}
Thomas~P. Minka.
\newblock Power {EP}.
\newblock Technical report, Microsoft Research, Cambridge, 2004.

\bibitem[Minka(2005)]{Minka:2005}
Thomas~P. Minka.
\newblock Divergence measures and message passing.
\newblock Technical report, Microsoft Research, Cambridge, 2005.

\bibitem[Minka and Lafferty(2002)]{Minka+Lafferty:2002}
Thomas~P. Minka and John Lafferty.
\newblock Expectation-propagation for the generative aspect model.
\newblock In \emph{Proceedings of the 18th Conference on Uncertainty in
  Artificial Intelligence (UAI-2002)}, pages 352--359. Morgan Kaufmann, San
  Francisco, CA, 2002.

\bibitem[Muirhead(2005)]{Muirhead:2005}
Robb~J. Muirhead.
\newblock \emph{Aspects of Multivariate Statistical Theory}.
\newblock John Wiley \& Sons, Hoboken, New Jersey, 2005.

\bibitem[Neiswanger et~al.(2014)Neiswanger, Wang, and
  Xing]{Neiswanger+others:2014}
Willie Neiswanger, Chong Wang, and Eric~P. Xing.
\newblock Asymptotically exact, embarrassingly parallel {MCMC}.
\newblock In \emph{Proceedings of the Thirtieth Conference on Uncertainty in
  Artificial Intelligence, {UAI} 2014, Quebec City, Quebec, Canada, July 23-27,
  2014}, pages 623--632, 2014.

\bibitem[Opper and Winther(2000)]{Opper+Winther:2000}
Manfred Opper and Ole Winther.
\newblock {Gaussian processes for classification: Mean-field algorithms}.
\newblock \emph{Neural Computation}, 12\penalty0 (11):\penalty0 2655--2684,
  2000.

\bibitem[Opper and Winther(2005)]{Opper+Winther:2005}
Manfred Opper and Ole Winther.
\newblock {Expectation consistent approximate inference}.
\newblock \emph{Journal of Machine Learning Research}, 6:\penalty0 2177--2204,
  2005.

\bibitem[Pearl(1986)]{pearl1986fusion}
Judea Pearl.
\newblock Fusion, propagation, and structuring in belief networks.
\newblock \emph{Artificial Intelligence}, 29\penalty0 (3):\penalty0 241--288,
  1986.

\bibitem[Ranganath et~al.(2016)Ranganath, Tran, and
  Blei]{ranganath2016hierarchical}
Rajesh Ranganath, Dustin Tran, and David~M. Blei.
\newblock Hierarchical variational models.
\newblock In \emph{International Conference on Machine Learning}, 2016.

\bibitem[Rasmussen and Williams(2006)]{Rasmussen+Williams:2006}
Carl~Edward Rasmussen and Christopher K.~I. Williams.
\newblock \emph{Gaussian Processes for Machine Learning}.
\newblock The MIT Press, 2006.

\bibitem[Riihim{\"a}ki et~al.(2013)Riihim{\"a}ki, Jyl{\"a}nki, and
  Vehtari]{Riihimaki+others:2013}
Jaakko Riihim{\"a}ki, Pasi Jyl{\"a}nki, and Aki Vehtari.
\newblock Nested expectation propagation for {Gaussian} process classification
  with a multinomial probit likelihood.
\newblock \emph{Journal of Machine Learning Research}, 14:\penalty0 75--109,
  2013.

\bibitem[Rue et~al.(2009)Rue, Martino, and Chopin]{Rue+others:2009}
H{\aa}vard Rue, Sara Martino, and Nicolas Chopin.
\newblock Approximate {Bayesian} inference for latent {Gaussian} models by
  using integrated nested {Laplace} approximations.
\newblock \emph{Journal of the Royal statistical Society B}, 71\penalty0
  (2):\penalty0 319--392, 2009.

\bibitem[Sahai(2018)]{Sahai:2018phd}
Swupnil Sahai.
\newblock \emph{Topics in Computational Bayesian Statistics With Applications
  to Hierarchical Models in Astronomy and Sociology}.
\newblock PhD thesis, Columbia University Academic Commons, 2018.

\bibitem[Sarr and Gupta(2009)]{Sarr+Gupta:2009}
Amadou Sarr and Arjun~K. Gupta.
\newblock Estimation of the precision matrix of multivariate {K}otz type model.
\newblock \emph{Journal of Multivariate Analysis}, 100\penalty0 (4):\penalty0
  742--752, 2009.

\bibitem[Sarwate et~al.(2014)Sarwate, Plis, Turner, Arbabshirani, and
  Calhoun]{Sarwate+others:2014}
Anand~D. Sarwate, Sergey~M. Plis, Jessica~A. Turner, Mohammad~R. Arbabshirani,
  and Vince~D. Calhoun.
\newblock Sharing privacy-sensitive access to neuroimaging and genetics data: A
  review and preliminary validation.
\newblock \emph{Frontiers in Neuroinformatics}, 8\penalty0 (35), 2014.
\newblock \doi{10.3389/fninf.2014.00035}.

\bibitem[Scott et~al.(2016)Scott, Blocker, and Bonassi]{Scott+others:2016}
Steven~L. Scott, Alexander~W. Blocker, and Fernando~V. Bonassi.
\newblock Bayes and big data: {The} consensus {Monte} {Carlo} algorithm.
\newblock \emph{International Journal of Management Science and Engineering
  Management}, 2016.
\newblock URL
  \url{http://www.tandfonline.com/doi/full/10.1080/17509653.2016.1142191}.

\bibitem[Seeger(2005)]{Seeger:2005}
Matthias Seeger.
\newblock Expectation propagation for exponential families.
\newblock Technical report, Max Planck Institute for Biological Cybernetics,
  Tubingen, 2005.

\bibitem[Seeger(2008)]{Seeger:2008}
Matthias Seeger.
\newblock Bayesian inference and optimal design for the sparse linear model.
\newblock \emph{Journal of Machine Learning Research}, 9:\penalty0 759--813,
  2008.

\bibitem[Seeger and Jordan(2004)]{Seeger+Jordan:2004}
Matthias Seeger and Michael~I. Jordan.
\newblock Sparse {G}aussian process classification with multiple classes.
\newblock Technical report, University of California, Berkeley, 2004.

\bibitem[Sivula(2015)]{Sivula:2015master}
Tuomas Sivula.
\newblock Distributed bayesian inference using expectation propagation.
\newblock Master's thesis, Aalto University, Espoo, Finland, 2015.

\bibitem[Smola et~al.(2004)Smola, Vishwanathan, and Eskin]{Smola+others:2004}
Alexander~J. Smola, Vishy Vishwanathan, and Eleazar Eskin.
\newblock {Laplace propagation}.
\newblock In S.~Thrun, L.K. Saul, and B.~Sch{\"o}lkopf, editors, \emph{Advances
  in Neural Information Processing 16}, 2004.

\bibitem[{Stan Development Team}(2017)]{Stan:2017}
{Stan Development Team}.
\newblock \emph{Stan modeling language: User's guide and reference manual},
  2017.
\newblock Version 2.17.0, \url{http://mc-stan.org/}.

\bibitem[Tran et~al.(2016)Tran, Ranganath, and Blei]{tran2016variational}
Dustin Tran, Rajesh Ranganath, and David~M. Blei.
\newblock The variational {Gaussian} process.
\newblock In \emph{International Conference on Learning Representations}, 2016.

\bibitem[Tsukuma and Konno(2006)]{Tsukuma+Konno:2006}
Hisayuki Tsukuma and Yoshihiko Konno.
\newblock On improved estimation of normal precision matrix and discriminant
  ccoefficients.
\newblock \emph{Journal of Multivariate Analysis}, 97\penalty0 (7):\penalty0
  1477--1500, 2006.

\bibitem[van Gerven et~al.(2009)van Gerven, Cseke, Oostenveld, and
  Heskes]{vanGerven+others:2009}
Marcel van Gerven, Botond Cseke, Robert Oostenveld, and Tom Heskes.
\newblock Bayesian source localization with the multivariate {Laplace} prior.
\newblock In Y.~Bengio, D.~Schuurmans, J.~Lafferty, C.~K.~I. Williams, and
  A.~Culotta, editors, \emph{Advances in Neural Information Processing 22},
  pages 1901--1909, 2009.

\bibitem[Vanhatalo et~al.(2013)Vanhatalo, Riihim{\"a}ki, Hartikainen,
  Jyl{\"a}nki, Tolvanen, and Vehtari]{Vanhatalo+others:2013}
Jarno Vanhatalo, Jaakko Riihim{\"a}ki, Jouni Hartikainen, Pasi Jyl{\"a}nki,
  Ville Tolvanen, and Aki Vehtari.
\newblock {GPstuff}: {Bayesian} modeling with {Gaussian} processes.
\newblock \emph{Journal of Machine Learning Research}, 14:\penalty0 1175--1179,
  2013.

\bibitem[Vehtari et~al.(2016)Vehtari, Gelman, and
  Gabry]{Vehtari+Gelman+Gabry:2016-PSIS}
Aki Vehtari, Andrew Gelman, and Jonah Gabry.
\newblock Pareto smoothed importance sampling.
\newblock \emph{arXiv:1507.02646}, 2016.

\bibitem[Villani and Larsson(2006)]{Villani+Larsson:2006}
Mattias Villani and Rolf Larsson.
\newblock The multivariate split normal distribution and asymmetric principal
  components analysis.
\newblock \emph{Communications in Statistics---Theory and Methods}, 35\penalty0
  (6):\penalty0 1123--1140, 2006.

\bibitem[Wand(2017)]{wand2017fast}
Matt~P. Wand.
\newblock Fast approximate inference for arbitrarily large semiparametric
  regression models via message passing.
\newblock \emph{Journal of the American Statistical Association}, 2017.

\bibitem[Wang and Dunson(2013)]{Wang+Dunson:2013}
Xiangyu Wang and David~B. Dunson.
\newblock Parallelizing {MCMC} via {W}eierstrass sampler.
\newblock \emph{arXiv preprint arXiv:1312.4605}, 2013.

\bibitem[Winn and Bishop(2005)]{Winn+Bishop:2005}
John Winn and Christopher~M. Bishop.
\newblock {Variational message passing}.
\newblock \emph{Journal of Machine Learning Research}, 6:\penalty0 661--694,
  2005.

\bibitem[Xu et~al.(2014)Xu, Lakshminarayanan, Teh, Zhu, and
  Zhang]{Xu+others:2014}
Minjie Xu, Balaji Lakshminarayanan, Yee~Whye Teh, Jun Zhu, and Bo~Zhang.
\newblock {Distributed {B}ayesian posterior sampling via moment sharing}.
\newblock In Zoubin Ghahramani, Max Welling, Corinna Cortes, Neil~D. Lawrence,
  and Kilian~Q. Weinberger, editors, \emph{Advances in Neural Information
  Processing Systems 27}, pages 3356--3364, 2014.

\bibitem[Zoeter and Heskes(2005)]{Zoeter+Heskes:2005}
Onno Zoeter and Tom Heskes.
\newblock {Gaussian quadrature based expectation propagation}.
\newblock In Robert Cowell and Zoubin Ghahramani, editors, \emph{International
  Workshop on Artificial Intelligence and Statistics (AISTATS)}, volume~10,
  2005.

\end{thebibliography}

\clearpage
\begin{appendices}

\section{Distributed parallel message passing algorithm}
\label{appendix_sec_algorithms}

This section presents a detailed algorithm for distributed EP applied in the context of partitioned data.
The implementation used in the experiments
in Section~\ref{experiments}
follows this description.
In Appendix~\ref{appendix_EP_dim_red}, we further extend the algorithm for cases where dimension reduction is possible.

\subsection{Algorithm description}\label{appendix_parallelep}

In this subsection we give a practical algorithm description suitable for
implementing the general message passing algorithms discussed
in Sections~\ref{general_framework} and~\ref{hier}.
The algorithm can be applied to approximate the joint posterior distribution in a general setting or the marginal posterior distribution of the shared parameters in a hierarchical setting.

Consider the normal distribution $g(\theta|r, Q) = \operatorname{N}(\theta | \mu, \Sigma)$ for the random variable $\theta \in \mathbb{R}^D$.
The precision mean vector $r \in \mathbb{R}^D$ and the symmetric (positive semidefinite) precision matrix $Q \in \mathbb{R}^{D \times D}$ are the natural parameters
and the mean vector $\mu \in \mathbb{R}^D$ and the symmetric (positive semidefinite) covariance matrix $\Sigma \in \mathbb{R}^{D \times D}$ are the moment parameters.
The parameters can be inverted from natural to moment form, $\Sigma = Q^{-1},\; \mu = Q^{-1}r$, and vice-versa, $Q = \Sigma^{-1},\; r = \Sigma^{-1}\mu$, using Cholesky factorization and backward substitution. Multiplying together two normal distributions yields an unnormalized normal distribution with natural parameters multiplied together
$g_1(\theta | r_1, Q_1) g_2(\theta | r_2, Q_2) \propto g_{1 \cdot 2}(\theta | r_1 + r_2, Q_1 + Q_2)$, and analogically $g_1(\theta | r_1, Q_1) / g_2(\theta | r_2, Q_2) \propto g_{1 / 2}(\theta | r_1 - r_2, Q_1 - Q_2)$ \citep[e.g.][p.\ 200]{Rasmussen+Williams:2006}.

EP is applied to approximate the target posterior distribution,
$$
  p(\theta|y) \propto p(\theta) \prod_{k=1}^K p(y_k| \theta),
$$
by a normal distribution,
$$
g(\theta | r, Q) \propto g_0(\theta | r_0, Q_0) \prod_{k=1}^K  g_k(\theta| r_k, Q_k),
$$
where the global approximation $g(\theta | r, Q)$, site approximations $g_k(\theta| r_k, Q_k)$, and the prior $g_0(\theta | r_0, Q_0)$ are all normal distributions parameterized by mean vector and precision matrix parameters. The global approximation parameters $r, Q$ can be obtained by summing up all the site parameters and the prior parameters:
$$
  Q = Q_0 + \sum_{k=1}^K Q_k, \qquad r = r_0 + \sum_{k=1}^K r_k.
$$

In the following algorithm description, the parameter $\eta \in (0,1]$ can be used to apply power EP~\citep{Minka:2004} to minimize general $\alpha$-divergence instead of KL divergence as discussed in
Section~\ref{approximation} in the paper.
Using $\eta=1$, as we did in our experiments, applies regular EP with KL divergence minimization.

Initially all the site distributions are set to improper uniform distributions with $r_k = 0, \; Q_k = 0$ for $k=1,2,\dots,K$,
which is equivalent to initializing the global approximation $g(\theta | r, Q)$ to the prior, that is, $r = r_0$ and $Q = Q_0$.
The algorithm proceeds by iteratively updating the site distributions until convergence:
\begin{enumerate}
    \setlength\itemsep{1em} % set some space between the items

\item In parallel at each site $k=1,2,\dots,K$,
determine the cavity distribution $g_{-k}(\theta | r_{-k}, Q_{-k})$:
$$
  Q_{-k} = Q - \eta Q_k, \qquad r_{-k} = r - \eta r_k.
$$
Here it is possible to obtain a precision matrix $Q_k$  that corresponds to an improper distribution (not in the first iteration). This is acceptable as long as the inference for the tilted distribution in the next step can be carried out. If this inference method requires a proper cavity distribution, the algorithm can jump to step~4, reduce damping, and continue until proper cavities are obtained.

\item In parallel at each site $k=1,2,\dots,K$,
approximate the natural precision parameters $r_{\setminus k}, Q_{\setminus k}$ of the tilted distribution
$$
g_{\setminus k}(\theta) \propto p(y_k | \theta)^\eta g_{-k}(\theta | r_{-k}, Q_{-k}),
$$
which is of unrestricted form. This can be sampled and differentiated using
$$
   \log g_{\setminus k}(\theta)
   = \eta \log p(y_k|\theta)
    -\frac{1}{2} \theta^T Q_{-k} \theta + r_{-k}^T \theta + \text{const}.
$$

In a hierarchical setting,
as discussed in Section~\ref{hier},
the tilted distribution considers also the local parameters:
$$
g_{\setminus k}(\theta) \propto
\int \bigl( p(y_k | \alpha_k, \theta) p(\alpha_k | \theta) \bigr)^\eta g_{-k}(\theta | r_{-k}, Q_{-k}) \:\mathrm{d}\alpha_k,
$$
where $\theta$ contains the shared parameters and $\alpha_k$ contains the local parameters for site $k$.

Key properties of different approximation methods are:
\begin{itemize}
  \item MCMC: It is easy to compute $\mu_{\setminus k}$ and
      $\Sigma_{\setminus k}$ from a set of simulation draws. Various approaches for
      computing the precision matrix $Q_{\setminus k} =
      \Sigma_{\setminus k}^{-1}$ are discussed in
      Appendix~\ref{appendix_invert_scatter} and
      in Section~\ref{app_precision}.
  \item Laplace's method: Gradient-based methods can be used to
      determine the mode of the tilted distribution
      efficiently. Once a local mode $\hat{\theta}$ is found,
      the natural parameters can be computed as
      \begin{align}
        Q_{\setminus k} &= -\nabla_{\theta}^2 \log g_{\setminus k}(\theta)
        |_{\theta=\hat{\theta}}
        = -\eta \nabla_{\theta}^2 \log p(y_k|\theta) |_{\theta=\hat{\theta}}
        + Q_{-k} \nonumber\\
        r_{\setminus k} &= Q_{\setminus k} \hat{\theta}. \nonumber
      \end{align}
      If $\hat{\theta}$ is a local mode, $Q_{\setminus k}$
      should be symmetric and positive definite.
\end{itemize}
Other approximation methods can also be used, including EP itself, which can be used to form arbitrarily deep and complex message passing algorithms.

The implementation used in the experiments
in Section~\ref{experiments}
uses MCMC sampling and consider the hierarchical structure of the problem.

\item In parallel at each site $k=1,2,\dots,K$,
if $|Q_{\setminus k}|>0$, compute the change in the site distribution $g_k(\theta | r_k, Q_k)$ resulting from the moment consistency conditions
$Q_{\setminus k} = Q_{-k} +\eta Q_{k}^{\rm new}$ and
$r_{\setminus k} = r_{-k} +\eta r_{k}^{\rm new}$:
\begin{align*}
  \Delta Q_{k} &= Q_{k}^{\rm new} - Q_k
  = \eta^{-1} (Q_{\setminus k} - Q_{-k}) -Q_k
  \\
  \Delta r_{k} &= r_{k}^{\rm new} - r_k
  = \eta^{-1} (r_{\setminus k} -r_{-k}) -r_k,
\end{align*}

If $|Q_{\setminus k}| \le 0$, there are at least two options:
discard the update by setting $\Delta Q_{k} = 0$ and $\Delta r_{k}
=0$, or use some method discussed
in Section~\ref{eigens}
to improve the conditioning of
$Q_{\setminus k}$ and compute the parameter updates with the
modified $Q_{\setminus k}$.

\item Update the global approximation $g(\theta | r, Q)$ with damping level $\delta \in (0,1]$:
\begin{align*}
Q^{\rm new} &= Q + \delta \sum_{k=1}^K \Delta Q_{k} \\
r^{\rm new} &= r + \delta \sum_{k=1}^K \Delta r_{k}.
\end{align*}
If the resulting approximation $g(\theta | r, Q)$ is not proper,
decrease $\delta$ and try again.

\item In parallel at each site $k=1,2,\dots,K$,
determine the updated site parameters with the selected damping level $\delta \in (0,1]$:
\begin{align*}
Q_{k}^{\rm new} &= Q_k + \delta \Delta Q_{k} \\
r_{k}^{\rm new} &= r_k + \delta \Delta r_{k}.
\end{align*}

\end{enumerate}
The iterations are repeated until all the tilted distributions are consistent
with the approximate posterior, that is, $\Delta r_k$ and $\Delta Q_k$ become small for all sites $k=1,2,\dots,K$.

\subsection{Advantages and limitations}

This section discusses advantages and limitations of the algorithm presented in Appendix~\ref{appendix_parallelep}.

\paragraph*{Advantages}
%Advantages of the approach include:
\begin{itemize}
\item Working with the natural parameters of the exponential family makes the computations in the algorithm convenient.
Operating with such terms can be parallelized elementwise, making the time complexity constant instead of $O(D^2)$, e.g.\ $[Q_1 + Q_2]_{i,j} = [Q_1]_{i,j} + [Q_2]_{i,j}$, where $[A]_{i,j}$ denotes element $i,j$ of matrix $A$.
Also, summing up multiple terms in step~4 can be parallelized termwise, e.g.\ $(Q_1 + Q_2) + (Q_3 + Q_4)$.
\item The tilted moments can be determined by sampling directly from the unnormalized tilted distributions or by using Laplace's method.
This requires only cheap function and gradient evaluations and can be applied to a wide variety of models.
\item After convergence, the final posterior approximation could be formed by mixing the draws from the different tilted distributions because these should be consistent with each other and with $g(\theta)$.
This sample-based approximation could also capture potential skewness in $p(\theta|y)$ because it resembles the EP-based marginal improvements described by \citet{Cseke+Heskes:2011}.
\end{itemize}

\paragraph*{Limitations}
%Limitations of the approach include:
\begin{itemize}
  \item The tilted distribution covariance matrices can be easily computed from the samples, but obtaining the precision matrix efficiently is problematic. Various methods for dealing with this issue are discussed
  in Section~\ref{app_precision}.
  These methods often involve computing the inverse of the sample covariance or scatter matrix, which as such is a costly and inaccurate operation. However, as discussed in Appendix~\ref{appendix_invert_scatter}, the QR-decomposition can be used here to more efficiently form the Cholesky factor of the matrix directly from the sample.

  \item Estimating the marginal likelihood is more challenging, because
      determining the normalization constants of the tilted distribution
      requires multivariate integrations. For example, annealed importance
      sampling type of approaches could be used if marginal likelihood estimates
      are required. We further discuss marginal likelihood in Appendix~\ref{appendix_marginal_likelihood}.

      With  Laplace's method, approximating the this normalization constant
      is straightforward but the quality of the marginal likelihood
      approximation is not likely to work well with skewed posterior
      distributions. The Laplace marginal likelihood estimate is not generally
      well-calibrated with the approximate predictive distributions in terms of
      hyperparameter estimation. Therefore it would make sense to integrate
      over the hyperparameters within the EP framework.
\end{itemize}

\subsection{Inverting the scatter matrix}\label{appendix_invert_scatter}

When using sample-based estimates for the tilted distribution moment estimation, one often needs to deal with the inverse of the scatter matrix (unnormalized sample covariance matrix). In practice, one wants to form the Cholesky decomposition for it. The naive way would be to calculate the scatter matrix and apply available routines to determine the factorization. However, here QR-decomposition can be used to compute it directly from the sample without ever forming the scatter matrix itself. This makes the process more stable, as forming the scatter matrix squares the condition number.

Consider the sample concatenated as an $n \times d$ matrix $D$ where the columns are centered to have zero mean. The scatter matrix is $S = D^T D$. In the QR-decomposition $D = QR$, the matrix $R$ corresponds to the upper triangular Cholesky factor of the scatter matrix, although the rows may be negative. Moreover, because the factor $Q$ is not needed, it is possible to compute the QR-decomposition even more efficiently.

\subsection{Estimating the natural parameters} \label{app_precision}

When using sample-based methods for the inference on the tilted distribution, one must consider the accuracy of the moment estimation.
In the message passing algorithm, these parameters are needed in natural form.
Estimating the moment parameters from a given sample is straightforward but estimating the precision matrix from a set of simulation draws is a complex task and in general the estimators are biased.
In addition, depending on the situation, using the naive unbiased moment estimator can produce higher expected KL divergence compared to other biased estimators.

In the algorithm, the parameters of the tilted distribution are needed in natural form.
The naive way of estimating the precision matrix $Q$ is to invert the unbiased sample covariance matrix, that is $\widehat{Q}=\widehat{\Sigma}^{-1} = (n-1)S^{-1}$, where $S$ is the scatter matrix constructed from the posterior simulation draws representing the tilted distribution. This estimator is biased in general:
%$\mathrm{E}\bigl(\widehat{Q\,|\,Q}\bigr) \neq Q$.
$\mathrm{E}\bigl(\widehat{Q}\bigr) \neq Q$.
Furthermore, the number of draws $n$ affects the accuracy of the estimate drastically. In an extreme case, when $n$ is less than the number of dimensions $d$, the sample covariance matrix is not even invertible as its rank can not be greater than $n$. In such a case, one could resort for example to the Moore–Penrose pseudo-inverse. In practice, when dealing with the inverse of the scatter matrix, one should apply the QR-decomposition to the samples in order to obtain the Cholesky decomposition of $S$ without ever forming the scatter matrix itself. This is discussed in more detail
in Appendix~\ref{appendix_invert_scatter}.

\begin{figure}[tb]
\centering
\input{figures/fig_kl_test.pgf}
\caption{%
Simulated histograms of resulting KL divergences $\operatorname{KL}(g_{\backslash k}(\theta)||g(\theta))$ when moment parameters are estimated using the naive unbiased moment estimator $S/(n-1)$ and the normal natural estimator $S/(n-d-2)$ presented in Equations~\eqref{eq:n_sample_prec_estim_Q} and~\eqref{eq:n_sample_prec_estim_r}.
The sample size for estimating the parameters $n=200$.
The tilted distribution is normally distributed with $d=16$ dimensions
and the correlation matrix is randomized with eigenvalues drawn from a Dirichlet distribution with a few higher values in the concentration parameter.
The sampling is repeated 8000 times to form a sample of the distribution of the KL divergences.
It can be seen from the histograms and from the illustrated statistics, that the biased estimator performs better in this case.
}\label{fig:kltest}
\end{figure}

\begin{figure}[tb]
\centering
\input{figures/fig_kl_test_scale.pgf}
\caption{%
Simulated statistics of resulting KL divergences $\operatorname{KL}(g_{\mathrm{EP}}(\theta)||g_{\text{sample}}(\theta))$, where $g_{\mathrm{EP}}(\theta)$ is the new global distribution with precise EP update and $g_{\text{sample}}(\theta)$ is the new global distribution with sample based update.
In the sample based update, different scaling of the scatter matrix is used, indicated in the x-axis as multiples of $1/n$.
The naive unbiased moment estimator $S/(n-1)$ and the normal natural estimator $S/(n-d-2)$ presented in Equations~\eqref{eq:n_sample_prec_estim_Q} and~\eqref{eq:n_sample_prec_estim_r}
are indicated by vertical lines.
The sample size for estimating the parameters $n=200$.
In the first row, the $d=16$ dimensional tilted distribution is normally distributed, and in the second row, it follows t-distribution with four degrees of freedom.
The correlation matrix is randomised in a similar fashion as in Figure~\ref{fig:kltest}.
The sampling is repeated 2000 times to form a sample of the distribution of the KL divergences for each estimator in a grid of 15 points.
It can be seen that while producing best results in the normal case, the normal natural estimator does not perform best in the t-distribution case.
}\label{fig:kltestscale}
\end{figure}

If the tilted distribution is normally distributed, an unbiased estimator for the precision matrix can be constructed by~\citet[][p. 136]{Muirhead:2005}:
\begin{align}
\widehat{Q}_\mathrm{N}
&= \frac{n-d-2}{n-1} \widehat{\Sigma}^{-1}
= (n - d - 2) S^{-1}.
\label{eq:n_sample_prec_estim_Q}
\intertext{Furthermore, the precision mean is given by,}
\widehat{r}_\mathrm{N}
&= \widehat{Q}_\mathrm{N} \widehat{\mu}
= (n - d - 2) S^{-1} \widehat{\mu},
\label{eq:n_sample_prec_estim_r}
\end{align}
which can be solved simultaneously while inverting the scatter matrix.
The inverse of this matrix is a biased estimate of the covariance matrix.
\citet{Xu+others:2014} used this estimator in their implementation of the MCMC based EP algorithm and conjectured that it would be an unbiased estimate of the true natural parameters of the tilted distribution, which is not true in general.
Other improved estimates for the normal distribution and some more general distribution families exist
\citep{Bodnar+Gupta:2011,Gupta+others:2013,Sarr+Gupta:2009,Tsukuma+Konno:2006}.
However, if the tilted distribution is normally distributed, it is likely that the moments can be solved analytically and sample based estimates are not needed in the first place.
Different methods for estimating the precision matrix in the general case, that is when no assumptions can be made about the tilted distribution, have also been proposed. These methods often either shrink the eigenvalues of the sample covariance matrix or impose sparse structure constraints to it
\citep{Bodnar+others:2014,Friedman+others:2008}.

In each iteration of EP, as discussed in Section~\ref{sec:ep}, the objective is to minimize the KL divergence from the global approximation to the tilted distribution, which corresponds to matching the moments.
With approximated moment estimates, in order to make the algorithm work like EP on expectation and stay at the same fixed point as EP on expectation, the moment estimates should be unbiased.
However, the distribution of the resulting KL divergence depends on the used moment parameter estimator.
An unbiased estimator is not necessarily optimal, as there can be biased estimators that produce smaller expected KL divergence.
The situation depends on the form of the tilted distribution and on the properties of the used estimators, e.g.\ the sample size and consistency in the case of simulation-based estimates.
Figures~\ref{fig:kltest} and~\ref{fig:kltestscale} illustrates a simulated example case, where the tilted distribution moments are estimated from a sample using various scatter matrix based estimators. Figure~\ref{fig:kltest} illustrates the distribution of the resulting KL divergence $\operatorname{KL}(g_{\backslash k}(\theta)||g(\theta))$ with normal tilted distribution $g_{\backslash k}(\theta)$ using the naive unbiased estimate and the normal distribution natural parameter estimates presented in Equations~\eqref{eq:n_sample_prec_estim_Q} and~\eqref{eq:n_sample_prec_estim_r}.
In this case, the biased moment estimator outperforms the unbiased naive one.
In Figure~\ref{fig:kltestscale}, a t distribution with four degrees of freedom is used.
In this case, the optimal scaling of the scatter matrix is greater than in both of the discussed estimators.
Thus, when implementing a simulation-based EP algorithm, it may be beneficial to study the form of the tilted distribution and select the used sample estimator accordingly.
The need for this selective analysis can be hindered by increasing the sample size;
With increased sample size, the naive moment and the normal natural estimators are likely to perform equally well.

\subsection{Dimension reduction for site inference}
\label{appendix_EP_dim_red}

The EP algorithm presented in Appendix~\ref{appendix_parallelep} can easily be extended to incorporate additional message passing components, as discussed, for example, by \citet{chen_wand2018factor_graph_fragments}.
Here we demonstrate a version for the special case in which the non-Gaussian likelihood terms $p(y_k|\theta)$ depend on $\theta$ only through preferably low-dimensional linearly transformed random variables $z_k = U_k^T \theta, \; U_k \in \mathbb{R}^{D \times D_k}, \; z_k \in \mathbb{R}^{D_k}$ for each partition $k=1,2,\dots,K$;
that is, $p(y_k|\theta) = p(y_k| z_k)$.

In the algorithm, the site approximations are stored in low-dimensional form (zero initialized) as $g_k(z_k | \widetilde{r}_k, \widetilde{Q}_k), \; \widetilde{r}_k \in \mathbb{R}^{D_k}, \, \widetilde{Q}_k \in \mathbb{R}^{D_k \times D_k}$. The global approximation $g(\theta | r, Q)$ in the original space can be obtained from the transformed site distributions and the prior distribution \citep[see, e.g.,][section Multivariate Linear Combination Derived Variable Fragment]{chen_wand2018factor_graph_fragments}
$$
  Q = Q_0 + \sum_{k=1}^K Q_k\quad
  r = r_0 + \sum_{k=1}^K r_k,
$$
where
\begin{equation}\label{appendix_eq_ep_transform_back}
Q_k = U_k \widetilde{Q}_k U_k^T, \qquad r_k = U_k \widetilde{r}_k.
\end{equation}
The algorithm proceeds similarly as in Appendix~\ref{appendix_parallelep}, but the site updates are performed in the transformed space:
\begin{itemize}
\item In step 1, the cavity distribution $g_{-k}(z_k | \widetilde{r}_{-k}, \widetilde{Q}_{-k})$ is calculated by,
$$
\widetilde{Q}_{-k} = \left( U_k^T Q^{-1} U_k \right)^{-1} - \eta \widetilde{Q}_k, \qquad
\widetilde{r}_{-k} = \left( U_k^T Q^{-1} U_k \right)^{-1} U_k^T Q^{-1} r - \eta \widetilde{r}_k.
$$
See, e.g., \cite[section Multivariate Linear Combination Derived Variable Fragment]{chen_wand2018factor_graph_fragments}.
\item In steps 2, 3, and 5, the computations are applied for
$z_k$, $\widetilde{r}_k$, and $\widetilde{Q}_k$ instead of
$\theta$, $r_k$, and $Q_k$.

\item In step 4, the global approximation is updated by transforming the difference into the original space by considering the relation in~\eqref{appendix_eq_ep_transform_back}:
\begin{align*}
Q^{\rm new} &= Q + \delta \sum_{k=1}^K \Delta Q_{k}
    = Q + \delta U_k \left[ \sum_{k=1}^K \Delta \widetilde{Q}_{k} \right] U_k^T\\
r^{\rm new} &= r + \delta \sum_{k=1}^K \Delta r_{k}
    = r + \delta U_k \left[ \sum_{k=1}^K \Delta \widetilde{r}_{k} \right].
\end{align*}

\end{itemize}

The advantage of the algorithm comes from the lower dimensional operation of the site updates; the tilted distribution inference considers only the transformed space of $z_k$ with $D_k$ dimensions instead of the space of $\theta$ with $D$ dimensions. In addition, the site distributions can be stored in the lower dimensional space with $O(D_k^2)$ elements in parameters $\widetilde{r}_k, \widetilde{Q}_k$ instead of the original space with $O(D^2)$ elements in parameters $r_k, Q_k$.

The disadvantage of the method is that the computation of the cavity distributions becomes heavier task with time complexity $O(D^3)$. However, if the global approximation moment parameters $\mu = Q^{-1}r\; \Sigma = Q^{-1}$ are solved between every iteration anyway, e.g.\ to monitor the convergence, the task does not add any complexity.

\section{Additional details of the experiments}\label{appendix:implementational_details}

As discussed in the text,
e.g.\ in Section~\ref{stuff},
implementing a distributed EP algorithm contains multiple design choices that might affect the behaviour of the algorithm. This section describes some details of the implementation we used in our experiments
in Section~\ref{experiments}.
In addition, we show some additional results obtained from the simulated experiment.
The algorithm follows the description
in Appendix~\ref{appendix_parallelep},
where the tilted distribution inference in step~2 considers the discussed hierarchical setting and is carried out by MCMC sampling.

\subsection{Implementation in Stan}\label{appendix_stan}

We implement our experiments using Python, R, and Stan.
The Python code for the simulated experiment is available at \url{https://github.com/gelman/ep-stan/releases/tag/v1.3}.
We pass the normal approximations $g_k$ back and forth between a master node and $K$ separate site nodes.
In the site nodes, we use Stan to compute the tilted distribution moments.

Our implementations are not optimal in methodological point-of-view. In the following, we list some key areas of improvement for our implementation:
\begin{itemize}
\item
In the current implementation, we write the appropriate Stan model for the tilted distribution inference manually by adapting the code from the full model to act just on the subset of parameters relevant to a single subset  of the partitioned hierarchical model.
In future software development, we would like to be able to take an existing Stan program and merely overlay a factorization so that the message passing algorithm could be applied directly.
\item
Currently, Stan performs adaptation each time it runs.
Future versions should allow restarting from the previous state, which should speed up computation substantially when the algorithm starts to converge.
\item
We should be able to approximate the expectations more efficiently using importance sampling.
\end{itemize}

\subsection{Simulated experiment marginal KL divergence}\label{appendix:ex1_extra_results}

In the simulated hierarchical logistic regression experiment
in Section~\ref{subsec:simulated_data_experiment},
the bottom subplot of Figure~\ref{fig:experiment_mse} shows the KL divergence from the reference posterior distribution to the approximate one. In addition to the mean and variance of the parameters, this measure also takes into account the correlations between the parameters.
Figure~\ref{appendix:fig:experiment_kld} features an analogous plot, where the correlation between the parameters is ignored; instead of the full KL divergence, the sum of the KL divergences between all the marginals is illustrated. With this measure, the difference in the accuracy between the EP approximation and the consensus MC approximation becomes bigger.

\begin{figure}[tb]
\centering
\input{figures/fig_ex1_timex_kld.pgf}
\caption{
Approximate marginal KL divergence of the posterior approximation from the target distribution as a function of the elapsed sampling time in the simulated hierarchical model.
Three methods are compared: full MCMC, distributed EP, and distributed consensus MC.
For EP (solid lines) and consensus MC (dotted lines), line colors indicate the number of partitions $K$.
The $y$-axis is in the logarithmic scale.
Compared to the full KL divergence illustrated in the analogous plot
in Figure~\ref{fig:experiment_mse},
the difference in the accuracy between EP and consensus MC is bigger.
}\label{appendix:fig:experiment_kld}
\end{figure}

\subsection{Details of the simulated hierarchical logistic regression}\label{appendix:ex1_details}

In the simulated hierarchical logistic regression problem
in Section~\ref{subsec:simulated_data_experiment},
the difficulty of the problem is controlled by regulating the resulting uncertainty in the simulated data. This is done by first fixing random model parameter values and then selecting suitable parameters for sampling the explanatory variable conditional to the model parameters for each group separately. Finally the response variable $y_{ij}$ is sampled given the explanatory variable $x_{ij}$ and group parameter $\beta_j$. The following describes these steps in detail.

First the hyperparameters are fixed. For the intercept, we set $\mu_0 = 1.5$ and $\log \sigma_0 = 0.4$, and for the slope, $\mu_d$ and $\log \sigma_d, d=1,2,\dots,D$ are drawn at random from $\operatorname{uniform}(-2,2)$ and $\operatorname{uniform}(-0.5, 0.5)$ respectively.
The prior for the hyperparameters is set so that $\mu_d \sim \operatorname{N}(0, 4^2)$ and $\mu_d \sim \operatorname{log-N}(0, 2^2)$ for $d=0,1,\dots,D$. Given the fixed hyperparameters, the group parameters $\beta_j$ are drawn at random according to the model distribution $\beta_{jd} \sim \operatorname{N}(\mu_d, \sigma_d^2)$.

The following describes the sampling of the data based on the fixed parameters $\beta_j$. Vectors (lowercase) and matrices (uppercase) are denoted with bold symbols to distinguish them from scalars. In addition, denoting the conditioning on the model parameters $\+\beta_j$ and the group indexing $j$ is omitted. Detailed derivations of the formulas are presented in Section~6.3.2 in~\citep{Sivula:2015master}.

The explanatory variable $\+x \in \mathbb{R}^D$ is sampled from normal distribution $\operatorname{N}(\+\mu, \sigma^2 \+\Sigma_0)$. The mean $\+\mu$ is restricted to equal in all dimensions: $\+\mu = \mu \mathds{1}$, where $\mathds{1}$ is a vector of 1's. The correlation structure $\+\Sigma_0$ is randomly generated using modified vines method by~\citet{lewandowski2009generating}, where the partial correlations are sampled from $\operatorname{Beta}(2,2)$ from the range $(-0.8, 0.8)$ and the diagonal is normalized to unity.

Consider the regression coefficient vector split into intercept coefficient $\alpha$ and slope $\+\beta$.
The classification uncertainty is controlled by setting restrictions to
$P = \operatorname{logit}^{-1}\bigl(\alpha + \+\beta^T \+x \bigr)$,
i.e.\ the probability of the response variable $y$ being in one class, which follows logit-normal distribution $\operatorname{logit-N}\bigl( \alpha + \+\beta^T \+\mu, \, \sigma^2 \+\beta^T \+\Sigma_0 \+\beta \bigr)$.
The resulting distribution of $P$ is restricted to have tail probabilities $\operatorname{Pr}(P \leq p_0) < \gamma_0$ and $\operatorname{Pr}(P > p_0) < \gamma_0$, where $p_0 = 0.2$ and $\gamma_0 = 0.01$.
In addition, smallest acceptable variance condition $\operatorname{Var}\bigl(\alpha + \+\beta^T \+x \bigr) \geq \tau_\mathrm{min}^2$ is set with $\tau_\mathrm{min} = 0.25$. Satisfying explanatory variable sampling parameters $\mu$ and $\sigma$ are then chosen by
\begin{align*}
\mu &= \begin{dcases}
    \frac{\delta_\text{max} -\alpha}{\sum_{d=1}^D \beta_i},
        & \text{if } \alpha > \delta_\text{max}, \\
    \frac{-\delta_\text{max} -\alpha}{\sum_{d=1}^D \beta_i},
        & \text{if } \alpha < -\delta_\text{max}, \\
    0   & \text{otherwise}.
    \end{dcases}
\\
\sigma &= \begin{dcases}
    \frac{\operatorname{logit}\left(p_0\right) + |\alpha|}
         {\operatorname{\Phi^{-1}}\left(\gamma_0\right) \sqrt{\+\beta^T \+\Sigma_0 \+\beta}},
        & \text{if } |\alpha| \leq \delta_\text{max}, \\
    \frac{\tau_\mathrm{min}}{\sqrt{\+\beta^T \+\Sigma_0 \+\beta}},
        & \text{otherwise},
    \end{dcases}
\end{align*}
where
$
\delta_\text{max} = \tau_\mathrm{min} \operatorname{\Phi^{-1}}\left( \gamma_0 \right) -\operatorname{logit}\left(p_0\right)
$
is the maximum magnitude of the mean of $\alpha + \+\beta^T \+x$. With this parameter selection method,
\begin{align*}
    P &\sim \text{logit-}\mathrm{N}\left( \alpha' , \, \frac{\operatorname{logit}\left(p_0\right) + \left| \alpha' \right|}
         { \operatorname{\Phi^{-1}}\left(\gamma_0\right)} \right),
\intertext{where}
    \alpha' &= \min\bigl( \max\left( \alpha, -\delta_{\text{max}} \right), \delta_{\text{max}}\bigr).
\end{align*}
{From} this, it can be seen that the resulting distribution of $P$ does not depend on $\+\beta$ or the dimensionality $D$. The parameter $\alpha$ tilts the distribution toward 0 or 1.

\section{The computational opportunity of parallel message passing algorithms}\label{appendix_cost}

We have claimed that message passing algorithms offer computational gains for large inference problems by splitting the data into pieces and performing inference on each of those pieces in parallel, occasionally sharing information between the pieces.  Here we detail those benefits specifically.

Consider the simple non-hierarchical implementation
in Section~\ref{general_framework}
with a multivariate normal approximating family, where the likelihood is factored into $K$ sites.
The tilted distribution is approximated with MCMC sampling.
Let $N_k$ be the number of data points in site $k$ and let $D$ be the number of parameters, that is, the length of the vector $\theta$.
We assume that we have $K+1$ parallel processing units: one central processor that maintains the global posterior approximation $g(\theta)$ and $K$ worker units on which inference can be computed on each of the $K$ sites.
The central unit stores the global approximation and the worker units store the respective site approximation.
Each distribution parameters consist of $O(D^2)$ values: mean or precision mean vector of length $D$ and covariance or precision matrix of size $D \times D$.
Furthermore, we assume a network transmission cost of $c$ per parameter.
Finally, we define $h(n,d)$ as the computational cost of generating a sample from a tilted distribution with $n$ data points and $d$ parameters.
In general case $h(n,d) \gg O(d^2 + n)$, where $O(d^2 + n)$ would be the minimal cost for analytically tractable case.

Each step of the algorithm then incurs the following costs:

\begin{enumerate}
\item {\bf Partitioning.}  This loading and caching step will in general have immaterial cost.

\item {\bf Initialization.}
The initialization of the approximations can be performed in parallel in every unit. Here it is assumed, that single parameter allocation is a constant time operation.

\item {\bf EP iteration.} Let $m$ be the number of iterations over all $K$ sites.  Empirically $m$ is typically a manageable quantity; however, numerical instabilities tend to increase this number.  In parallel EP, damped updates are often used to avoid oscillation~\citep{vanGerven+others:2009}.

\begin{enumerate}

\item Computing the cavity distribution.
First, the current global approximation needs to be sent from the master node to the worker nodes with cost $O(cKD^2)$. On the worker nodes in parallel, this step involves only simple subtraction of $O(D^2)$ values per site, which can be parallelized locally. Thus the resulting total cost is $O(cKD^2)$.

% \item Forming the tilted distribution.  This conceptual step bears no cost.

\item Fitting an updated local approximation $g_k(\theta)$.
This step is performed on parallel in every worker. First a sample from the tilted distribution is generated with cost $h(N_k,D)$. After this, moment estimates are generated in natural form based on the obtained sample. As discussed in Appendix~\ref{appendix_invert_scatter},
QR-decomposition is used to form the estimates with cost $O(D^3)$. The resulting total cost is $O(h\left(\max N_k,D \right) + D^3)$.

\item Return the updated $g_k(\theta)$ to the central unit. This cost repeats the cost and consideration of step~3a.

\item Update the global approximation $g(\theta)$. Summing up the updates from all the sites has the naive cost of $O(KD^2)$, or improved cost of $O(\log K)$, when summations are parallelized elementwise and termwise. However, if the cost $h$ of approximating the posterior distribution is variable across worker units, the central unit could update $g(\theta)$ whenever possible while waiting for other sites to finish.

\end{enumerate}

\end{enumerate}

Considering only the dominating terms, across all these steps and the $m$ EP iterations, we have the total cost of our parallel message passing algorithm:
\begin{equation*}
O\left( m \left(cKD^2 + h\left(\max N_k,D \right) + D^3 \right) \right).
\end{equation*}
By comparison, consider first the cost of a non-parallel version:
\begin{equation*}
O\left( m \left(KD^3 + \textstyle{\sum_k} h\left(N_k,D \right) \right) \right).
\end{equation*}
Second, consider the cost of full sampling with no partitioning:
\begin{equation*}
O\left(h\Bigl(\textstyle{\sum_k} N_k,D \Bigr)\right).
\end{equation*}
With these three expressions, we can immediately see the computational benefits of our scheme.
In many cases, sampling will be by far the most costly operation, and will depend superlinearly on its arguments.
Thus, the parallel message passing scheme will dominate.  As the total data size $N=\sum_k N_k$ grows large, our scheme becomes essential.
When data is particularly big (e.g.\ $N \approx 10^9$), our scheme will dominate even in the rare case that $h(n,d)$ is in its minimal $O(d^2 + n)$.

\section{Comparison of SNEP and moment matching}\label{appendix_snep_vs_ep}

Section~\ref{approximation} discussed the SNEP method introduced by \citet{Hasenclever+others:2017}, which can be used for a simulation-based site inference instead of the tilted distribution moment matching method. In this section, we shortly compare both methods in two experiments. The experiments do not give an exhaustive view on the difference of the behaviour of the methods in all situations. However, they show that both of the methods have pros and cons and that the preferability of the methods is situational. The source code for both experiments is available online at \url{https://github.com/gelman/ep-stan}.

\begin{figure}[tb]
\centering
   \includegraphics[width=\textwidth]{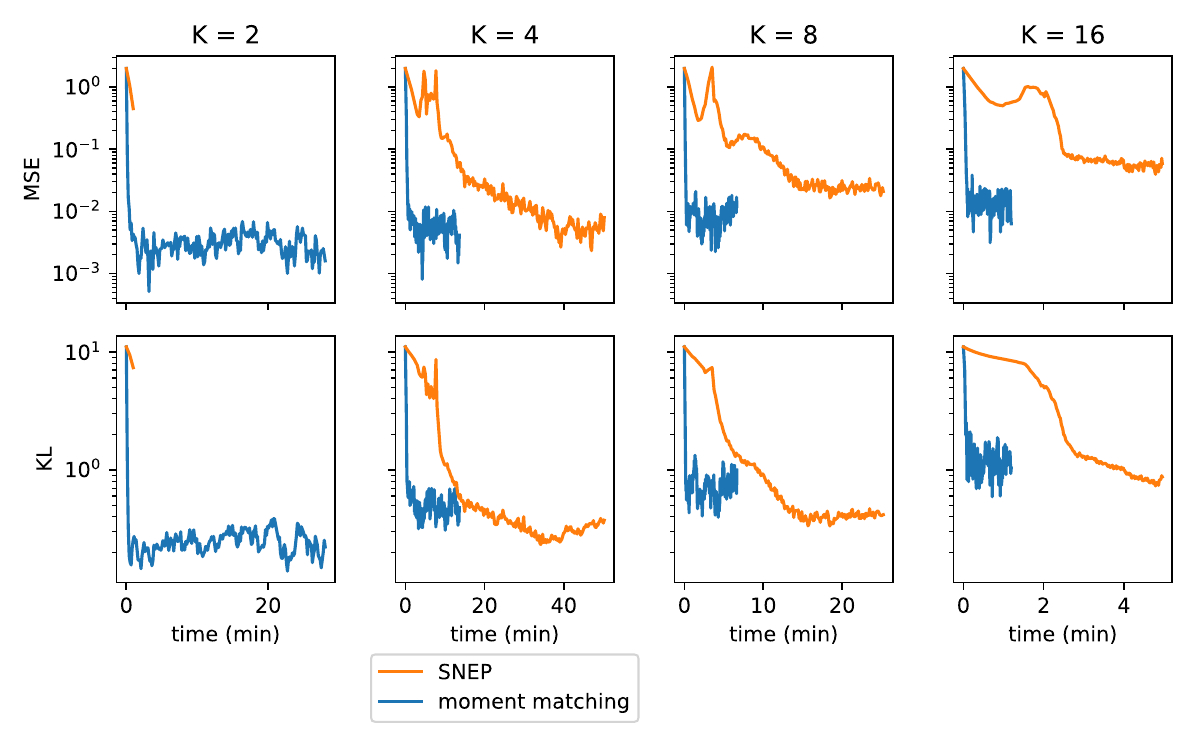}
\caption{
Comparison of simulation-based moment matching EP and SNEP in a simple hierarchical logistic regression experiment. The plots show the mean squared error of the mean and approximate Kullback-Leibler divergence from the target distribution to the resulting posterior approximation as a function of the elapsed sampling time. The $y$-axis is in the logarithmic scale. Each column corresponds to a different number of partitions $K$. When $K=2$, SNEP results in an improper global approximation and fails after two iterations. In other runs, both methods converge to a solution with comparable accuracy. SNEP reaches the solution slower and has more variability in the beginning. Moment matching has a faster start but more variability in the end.
}\label{fig:app_snep_ep_timex}
\end{figure}

We replicate the simulated hierarchical logistic regression experiment on Section~\ref{subsec:simulated_data_experiment} in a smaller scale with $J=16$ hierarchical groups and $D=3$ dimensions of the explanatory variable. The problem is distributed into $K=2,4,8,16$ sites. We apply a constant damping factor of 0.8 for SNEP and 0.5 for moment matching. We apply four inner iterations with auxiliary parameter updating in every other iteration in the site update for SNEP.
Because SNEP is not compatible with uniform initial site distributions, these distributions were set to
$\mathrm{N}(0, 2 K \max(\operatorname{Cov}(p(\theta))) \mathrm{I})$
instead.
All other settings were identical to the bigger experiment of Section~\ref{subsec:simulated_data_experiment}.
The results of the simulation are illustrated in Figure~\ref{fig:app_snep_ep_timex}, which shows the accuracy of the approximation as a function of the elapsed sampling time. As with  Figure~\ref{fig:experiment_mse} for the bigger experiment in Section~\ref{subsec:simulated_data_experiment}, in order to focus on the significant portion of the algorithm and ignore some implementation dependent factors, we do not in this graph count time spent in non-sampling parts of the algorithm. Compared to the moment matching method, the site update portion of SNEP contains more heavy operations, such as Cholesky factorisations of order $d_\phi$, and consequently SNEP spent more time in non-sampling parts of the algorithm in our experiment.
When the problem was distributed to $K=2$ sites, SNEP algorithm failed in all sites after two iterations by resulting in an improper global approximation and the iteration had to be terminated.
In other runs, both methods converge to a relatively similar solution but not into identical one. Moment matching converged faster than SNEP but had more variability in the end. SNEP was slower and had more variability in the start.

\begin{figure}[tbp]
\centering
\includegraphics[width=0.85\textwidth]{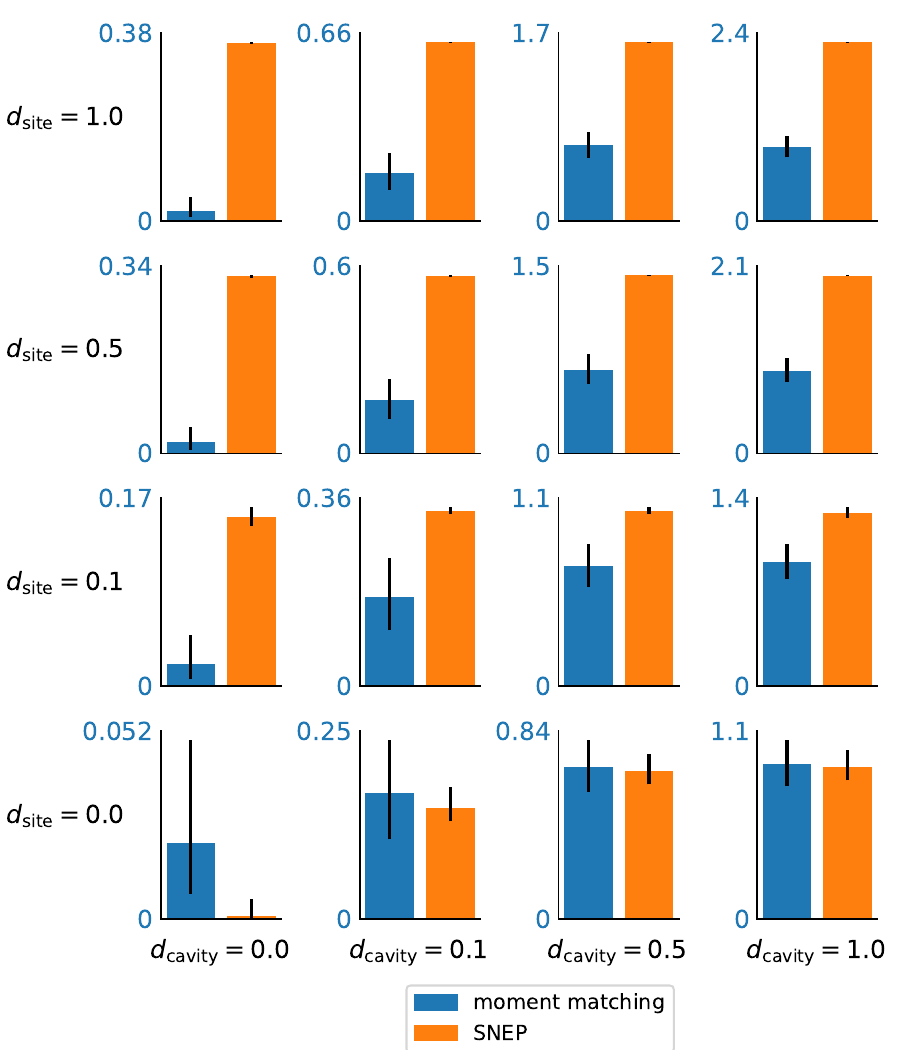}
\caption{
KL divergence from the target distribution to the resulting posterior approximation after updating one site with different initial site settings.
In each row and column, $d_\text{site}$ and $d_\text{cavity}$ indicate the divergence of the respective initial distribution from the convergence, where $0.0$ corresponds to the converged state and $1.0$ to a diverged state.
In order to focus on the difference between the methods, the y-axis is in different scale in each tile.
The simulation is repeated 2000 times. The bar heights show the medians of the obtained resulting KL divergences, and the attached black lines indicate 2.5\% and 97.5\% quantiles.
It can be seen from the figure that moment matching method has greater variability in general. In this case, moment matching outperforms SNEP when  thesite distribution is not completely converged.
}\label{fig:app_snep_ep}
\end{figure}

In addition to the single-run simulations, we also set up an experiment for testing the effect of single site updating in different settings.
We set up a randomized three-dimensional normal target distribution consisting of six normal factorized parts with variances ranging from $\mathrm{e}$ to $\mathrm{e^{-1}}$ and a normal prior distribution $p(\theta) \sim \mathrm{N}(0, 4^2 \mathrm{e I})$. Each factor is approximated in one site. We measure the resulting global approximation KL divergence after performing one site update with different initial site and cavity distributions. These distributions are set
\begin{gather*}
    g_k(\theta) = \Big( f_k(\theta) \Big)^{1 - d_\text{site}}
        \left(g_k^\text{extreme}(\theta)\right)^{d_\text{site}} \\
    g_{-k}(\theta) = \left(\prod_{i \neq k} f_i(\theta)\right)^{1 - d_\text{cavity}}
        \left(g_{-k}^\text{extreme}(\theta)\right)^{d_\text{cavity}},
\end{gather*}
where $d_\text{site}$ and $d_\text{cavity}$ control the deviation from the converged state,
$f_k(\theta)$ is the true underlying factor and the site distribution at the convergence, and
$\prod_{i \neq k} f_i(\theta)$ is the cavity distribution at the convergence.
The extreme site and cavity distributions are set to $g_k^\text{extreme}(\theta) \sim \mathrm{N}(0, 2^2 \mathrm{e I})$ and $g_{-k}^\text{extreme}(\theta) = \left(g_k^\text{extreme}(\theta)\right)^{K-1} p(\theta)$ respectively.
For both methods, the sample size at the site update is 200. For SNEP, we apply one inner iteration for the site update.
We replicate the experiment 2000 times.
The results of the experiment are illustrated in Figure~\ref{fig:app_snep_ep}.
Experimenting with four inner SNEP iterations with auxiliary parameter update in every other iteration did not change the results much.

Based on the discussed two experiments, it is clear that the methods behave differently and have characteristics that make them better in different situations.
The experiments suggest that SNEP behaves more chaotically and progresses slower when far away from the convergence. On the other hand, it seems to be more stable when sufficiently close to the convergence.
We believe this behavior is related to SNEP operating on the moment domain instead of the natural domain as it is done in moment matching.
To the best of our knowledge, this feature makes SNEP also incompatible with uniform initial site distributions.
One interesting idea would be to apply moment matching in early iterations for possibly more stable and faster start and switch to using SNEP for later iterations for more stable convergence.
Further study is needed in order to draw more elaborate conclusions about their differences.

\section{Marginal likelihood}\label{appendix_marginal_likelihood}

Although not the focus of this work, we mention in passing that EP also offers as no extra cost an approximation of the marginal likelihood, $p(y)=\int \! p_0(\theta) p(y|\theta)\,\mathrm{d}\theta$. This quantity is often used in model choice.

To this end, associate a constant $Z_k$ to each approximating site $g_k(\theta)$ and write the global approximation as,
$$ g(\theta) = p_0(\theta) \prod_{k=1}^K \frac{1}{Z_k} g_k(\theta).$$

Consider the Gaussian case, for the sake of simplicity, so that
$ g_k(\theta) = e^{ -\frac{1}{2}\theta^T Q_k\theta + r_k^T\theta}$,
under natural parameterization, and denote by $\Psi(r_k,Q_k)$ the corresponding normalizing constant:
$$ \psi(r_k,Q_k) = \int  e^{-\frac{1}{2}\theta^T Q_k\theta + r_k^T\theta}\,\mathrm{d}\theta
= \frac{1}{2}(-\log|Q_k/2\pi|+ r_k^T Q_kr_k).
$$
Simple calculations~\citep{Seeger:2005} then lead to following formula for the update of
$Z_k$ at site $k$:
$$ \log(Z_k) = \log(Z_{\setminus k}) -\Psi(r,Q) + \Psi(r_{-k},Q_{-k}),
$$
where $Z_{\setminus k}$ is the normalizing constant of the tilted distribution $g_{\setminus k}(\theta)$; $(r,Q)$ is the natural parameter of $g(\theta)$; and
$r\!=\!\sum_{k=1}^K \! r_k$, $Q\!=\!\sum_{k=1}^K \! Q_k$, $r_{-k}\!=\!\sum_{j\neq k} r_j$,
$Q_{-k}\!=\!\sum_{j\neq k} Q_j$. For deterministic approaches, we have discussed
approximating the moments of $g_{\setminus k}(\theta)$, it is straightforward to obtain an approximation of the normalizing constant; when simulation is used, some extra efforts may be required, as in \citet{Chib:1995}.

Finally, after completion of EP, one should return the quantity,
$$ \sum_{k=1}^K \log(Z_k) + \Psi(r,Q) -\Psi(r_0,Q_0),
$$
as the EP approximation of $\log p(y)$, where $(r_0,Q_0)$ is the natural parameter of the prior.

\section{Optimizing EP energy}\label{appendix_ep_energy}

Consider an EP approximation in the exponential family distribution,
\begin{gather*}
    g(\theta) = \frac{1}{Z} p(\theta)\prod_{k=1}^n g_k(\theta),
\end{gather*}
where $Z$ is a normalization constant.
The global approximation can be formulated as $g(\theta) \propto p(\theta) \exp(\+s^T \+\lambda)$ and the cavity distribution as $g_{\setminus k}(\theta) \propto p(\theta) \exp(\+s^T \+\lambda_{\setminus k})$, where $\+\lambda$ and $\+\lambda_{\setminus k}$ denote natural parameters.
A fixed point of the EP algorithm corresponds to a stationary point of the following objective function~\citep{Minka:2001b}:
\begin{gather*}
    \min_\lambda \max_{\lambda_{\setminus k}} \;
    (K-1) \log \int\! p(\theta) \exp(\+s^T \+\lambda) \,\mathrm{d}\theta
    - \sum_{k=1}^K \log \int \!p(\theta) p(y_k | \theta) \exp(\+s^T \lambda_{\setminus k}) \,\mathrm{d}\theta \\
    \text{such that} (K-1) \lambda = \sum_{k=1}^K \lambda_{\setminus k}.
\end{gather*}
This objective function corresponds to $-\log Z$ and to the expectation-consistent approximation~\citep{Opper+Winther:2005}.
The correspondence and connection to the Bethe free energy is demonstrated by~\citet{heskes2005}.

\end{appendices}

\end{document}